\documentclass[showpacs,amsmath,amssymb,twocolumn,superscriptaddress,notitlepage,preprintnumbers,pra]{revtex4-2}

\usepackage[dvips]{graphicx}
\usepackage{amsmath,amssymb,amsthm,mathrsfs,amsfonts,dsfont}

\usepackage{dcolumn}
\usepackage{bm}
\usepackage{amsmath}
\usepackage{amssymb}
\usepackage{braket}
\usepackage{physics}
\usepackage{amsthm}
\usepackage{natbib}
\usepackage{mathtools}
\usepackage{diagbox}
\usepackage{hyperref}
\usepackage[utf8]{inputenc}
\usepackage[english]{babel}
\usepackage{scalerel}[2014/03/10]
\usepackage{stackengine}
\usepackage{algorithm}
\usepackage{relsize}
\usepackage[noend]{algpseudocode} 
\usepackage[capitalize]{cleveref}
\usepackage{hyperref}
\hypersetup{colorlinks=true, linkcolor=blue, citecolor=blue, urlcolor=blue }

\newcommand{\degree}{^\circ}

\newcommand{\SDU}{School of Physics, State Key Laboratory of Crystal Materials, Shandong University, Jinan 250100, China}

\begin{document}


\title{Measurement-Device-Independent Entanglement Quantification in a Fully Connected Time-Bin Quantum Network}

\author{Lu Liu}
\email{These authors contributed equally to this work.}
\affiliation{\SDU}
\author{Ling-Xuan Kong}
\email{These authors contributed equally to this work.}
\affiliation{\SDU}
\author{Ze-Yang Lu}
\affiliation{\SDU}
\author{Xu-Jie Peng}
\affiliation{\SDU}
\author{Xiao-Xu Fang}
\email{fangxiaoxu@sdu.edu.cn}
\affiliation{\SDU}
\author{He Lu}
\email{luhe@sdu.edu.cn}
\affiliation{\SDU}

\begin{abstract}
Fully connected quantum networks enable scalable quantum communication, yet reliable entanglement characterization without trusting measurement devices remains challenging. Here we experimentally demonstrate measurement-device-independent (MDI) entanglement verification and quantification in a time-bin–encoded fully connected quantum network. Using a broadband periodically poled lithium niobate on insulator source combined with dense wavelength-division multiplexing, we distribute all six pairwise entangled links among four users over 20-km fiber channels, preserving high-fidelity entanglement without active stabilization of the long-distance fiber links. We show that conventional entanglement witnesses can fail under untrusted measurement conditions. By encoding trusted input states in the polarization degree of freedom of the same photons, we realize MDI measurements without ancillary photons or additional experimental resources. Both entanglement verification and quantification are obtained from the same measurement dataset. Our results establish a practical and scalable approach for reliable entanglement characterization in quantum networks.
\end{abstract}

\maketitle

Fully connected quantum networks (FCQNs), in which entanglement is shared between every pair of users, provide a powerful architecture for scalable quantum communication and quantum key distribution~(QKD)~\cite{Wengerowsky2018Nature}. Wavelength multiplexing enables multiple users to share a common broadband entangled photon source, significantly reducing resource overhead compared to point-to-point links. Rapid experimental progress has demonstrated FCQNs based on both polarization and time-bin encoding~\cite{Wengerowsky2018Nature, Joshi2020Sci.Adv., Alshowkan2021PRXQ,Liu2022PhotoniX, Shi2025arXiv, Huang2026NP, Kim2022APL.Photon., Fitzke2022PRXQ, Huang2025LPR}, with further scalability enabled by techniques such as entanglement swapping~\cite{Li2019PRL,Huang2026NP}. Among these approaches, time-bin encoding is particularly well suited for fiber-based implementations due to its intrinsic robustness against polarization fluctuations and environmental noise, making it highly attractive for long-distance quantum networks~\cite{Li2019PRL, Kim2022APL.Photon., Fitzke2022PRXQ, Huang2025LPR, Xavier2025npjQI}.

As quantum networks grow in size and complexity, reliable verification and quantification of distributed entanglement become central challenges. Conventional approaches, such as quantum state tomography and entanglement witnesses~\cite{james2001PRA, thew2002PRA,horodecki1996PLA,terhal2000PLA, guhne2009PR}, rely on the accurate characterization of measurement devices~\cite{rosset2012PRA,sarkar2026NP,morelli2022PRL, cao2024PRL}. These assumptions are difficult to satisfy in large-scale and interferometer-based systems, particularly for time-bin encoding, where phase stability and precise interferometric control are required at multiple nodes. The measurement-device-independent~(MDI) framework addresses this limitation by removing the need to trust measurement devices~\cite{buscemi2012PRL, branciard2013PRL}. MDI entanglement witnesses~(MDI-EW) have been experimentally demonstrated in bipartite systems~\cite{xu2014PRL, verbanis2016PRL} and extended to multipartite~\cite{li2020PRL}, continuous-variable~\cite{abiuso2021PRL, wang2025PRL}, and network scenarios~\cite{fu2025LSA}. However, existing implementations remain difficult to scale and primarily focus on entanglement certification rather than quantification. In particular, many approaches rely on auxiliary photons as trusted inputs, introducing significant resource overhead, while experimentally accessible entanglement quantification within the MDI framework remains largely unexplored.

\begin{figure*}[htp]
\centering
\includegraphics[width=\textwidth]{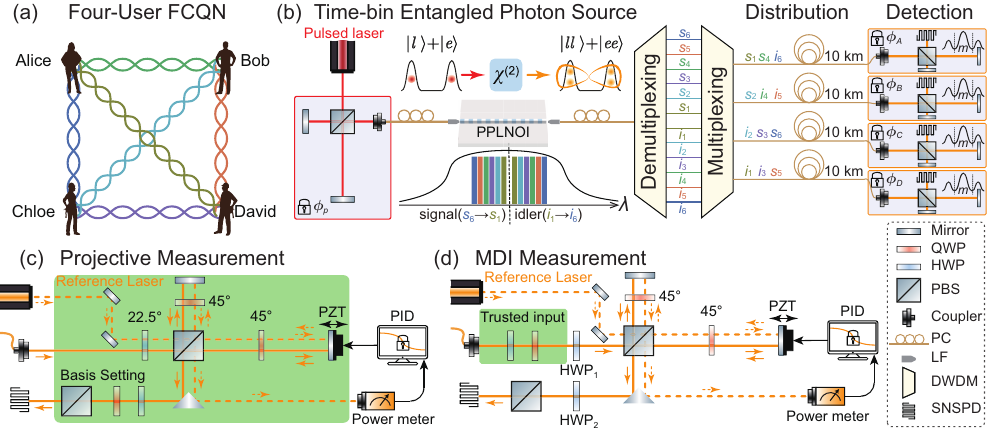}
\caption{\textbf{Schematic illustration of the setups of the four-user FCQN and its characterization. (a)}~Conceptual schematic of the FCQN, where four users (Alice, Bob, Chloe, and David) share all six pairwise entangled links. \textbf{(b)}~Time-bin entangled photon-pair source based on a PPLNOI waveguide. A pulsed laser is prepared in a time-bin superposition via an unbalanced Mach–Zehnder interferometer~(UMZI) and drives broadband spontaneous parametric down-conversion~(SPDC). The generated photon pairs are wavelength-demultiplexed and distributed to four users via dense wavelength-division multiplexing~(DWDM). \textbf{(c)}~Experimental setup for conventional projective measurements, where time-bin states are converted into polarization states using UMZI with phase stabilization. \textbf{(d)}~Experimental setup for the measurement-device-independent~(MDI) measurement, where trusted quantum inputs are encoded in the polarization degree of freedom~(DoF) and Bell-state measurements~(BSMs) are performed between the polarization and time-bin DoFs of the same photon. The green shaded region indicates the devices that are required to be well characterized in these schemes. Symbols: QWP, quarter-wave plate; HWP, half-wave plate; PBS, polarizing beam splitter; PZT, piezoelectric transducer; PC, polarization controller; LF, lensed fiber; DWDM, dense wavelength-division multiplexer; SNSPD, superconducting nanowire single-photon detector.}
\label{fig:setup}
\end{figure*}

Here, we experimentally demonstrate MDI entanglement verification and quantification in a time-bin–encoded FCQN. Using a broadband periodically poled lithium niobate on insulator~(PPLNOI) source combined with dense wavelength-division multiplexing~(DWDM), we distribute all six pairwise entangled links among four users over 20-km fiber channels. We show that conventional entanglement witnesses can fail under untrusted measurement conditions, falsely certifying separable states as entangled. MDI measurements are realized by encoding trusted input states in the polarization degree of freedom~(DoF) of the same photons, eliminating the need for ancillary photons and additional experimental resources. Within this framework, entanglement verification and quantification are obtained from the same measurement dataset. Our results provide a practical and scalable approach for reliable entanglement characterization in FCQNs.

We experimentally realize a four-user FCQN using a wavelength-multiplexed time-bin entangled photon source. As illustrated in Fig.~\ref{fig:setup}~(a), this is achieved by distributing six pairwise entangled links among Alice ($A$), Bob ($B$), Chloe ($C$), and David ($D$), corresponding to the connections $AB$, $AC$, $AD$, $BC$, $BD$, and $CD$, each ideally prepared in the maximally entangled state $\ket{\Phi^+}=(\ket{00}+\ket{11})/\sqrt{2}$. We label the users by $u,v\in\{A,B,C,D\}$ and denote each link by $(u,v)$. The network is implemented using a broadband time-bin entangled photon-pair source based on a PPLNOI waveguide. As shown in Fig.~\ref{fig:setup}~(b), a pulsed pump laser at 775~nm (100~MHz repetition rate, $\sim$10~ps pulse duration) is prepared in a coherent superposition of early~($\ket{e}$) and late~($\ket{l}$) temporal modes using an unbalanced Mach–Zehnder interferometer~(UMZI), yielding the state $\frac{1}{\sqrt{2}}(\ket{e_p}+\ket{l_p})\otimes\ket{V}$, where $\ket{H}$ ($\ket{V}$) denotes horizontal (vertical) polarization. The two time bins are separated by $\sim 640~\mathrm{ps}$, much larger than the pulse duration, ensuring that they are well resolved. This pump drives type-0 spontaneous parametric down-conversion~(SPDC) in the PPLNOI waveguide, generating photon pairs in the state $\frac{1}{\sqrt{2}}(\ket{ee}+\ket{ll})\otimes\ket{VV}$. The broad SPDC bandwidth (3-dB bandwidth $\sim 1126~\mathrm{nm}$) enables efficient wavelength multiplexing in the telecom regime.

Using DWDM, the broadband photon pair is separated into multiple frequency-correlated signal–idler channel pairs $(s_j,i_j)$. In the experiment, six channel pairs are selected and distributed to the four users, thereby establishing all pairwise entangled links simultaneously. For each link $(u,v)$, the two photons are transmitted through separate 10-km fiber channels, corresponding to a total separation of 20-km. This wavelength-multiplexed scheme enables the simultaneous realization of a FCQN from a single integrated source. Further details are provided in the Supplementary Materials.

\begin{figure}[htbp]
\centering
\includegraphics[width=\linewidth]{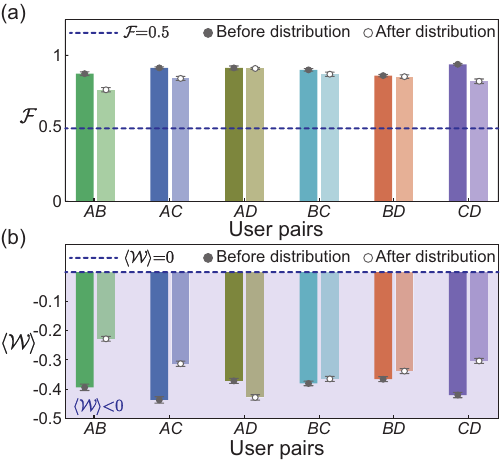}
\caption{\textbf{Characterization of the distributed entangled states for all six links in the FCQN, before and after 20-km fiber distribution. (a)}~Tomographic fidelities of the reconstructed two-photon states for each user pair. Gray (white) dots correspond to results obtained before (after) fiber distribution. The dashed line at $\mathcal{F}=0.5$ indicates the classical limit~\cite{Terhal2000PRA,guhne2009PR}. \textbf{(b)}~Measured entanglement witness values for the same links. The dashed line at $\langle \mathcal{W} \rangle = 0$ marks the entanglement bound. }
\label{fig:tomo_fidelity}
\end{figure}

To characterize the constructed FCQN, we first perform quantum state tomography based on local projective measurements. As shown in Fig.~\ref{fig:setup}~(c), each user employs an UMZI with a path-length difference matched to that used in the preparation of pump light, ensuring temporal overlap between the early–long and late–short components of the time-bin qubit. Under this condition, these components become indistinguishable and interfere coherently, effectively mapping the time-bin information onto the polarization DoF
\begin{equation}
\begin{split}
&\frac{1}{\sqrt{2}}\left(\ket{e_ue_v}+\ket{l_ul_v}\right)\otimes\ket{V_uV_v}
\\
\xrightarrow{\mathrm{UMZI}}&
\frac{1}{\sqrt{2}}\left(\ket{H_uH_v}+\ket{V_uV_v}\right)\otimes\ket{m_um_v}
\end{split}
\end{equation}
By postselecting events in the middle time bins $\ket{m_u m_v}$, the state is converted into a polarization-entangled state, enabling standard polarization analysis for tomography using half-wave plates~(HWP), quarter-wave plates~(QWP), and polarizing beam splitter~(PBS).

For each link $(u,v)$, the two-photon density matrix $\rho_{uv}$ is reconstructed before and after transmission through the 20-km fiber link. The fidelities with respect to the maximal entangled state $\ket{\Phi^+}=(\ket{HH}+\ket{VV})/\sqrt{2}$ is calculated by $\mathcal F=\mathrm{Tr}(\ket{\Phi^+}\bra{\Phi^+}\rho_{u,v})$, and the results are shown in Fig.~\ref{fig:tomo_fidelity}~(a). High fidelities are observed for all links, ranging from $0.86\pm0.01$ to $0.94\pm0.01$ (average $0.90\pm0.01$) before transmission and from $0.76\pm0.01$ to $0.91\pm0.01$ (average $0.84\pm0.01$) after transmission. Notably, no active stabilization or compensation is applied during fiber transmission. Phase stabilization is required only locally at the analysis UMZIs. The preservation of high fidelities across all links thus highlights the intrinsic robustness of time-bin encoding against polarization fluctuations and environmental perturbations in optical fibers. We further characterize the distributed links using entanglement witness, which enables direct certification of entanglement from a reduced set of measurements. For the target Bell state $\ket{\Phi^+}$, we employ the witness $\mathcal W=\mathbb I_4/2-\ket{\Phi^+}\bra{\Phi^+}$~\cite{Horodecki1999PRA, guhne2009PR}, equivalently written as 
\begin{equation}\label{Eq:EW}
    \mathcal W=\frac{1}{4}\left(\mathbb I_2\otimes \mathbb I_2-\sigma_x\otimes\sigma_x+\sigma_y\otimes\sigma_y-\sigma_z\otimes\sigma_z\right),
\end{equation}
where $\sigma_x$, $\sigma_y$, and $\sigma_z$ denote the Pauli operators and $\mathbb I$ is the identity. The measured witness values, shown in Fig.~\ref{fig:tomo_fidelity}~(b), are negative for all links, confirming entanglement across all user pairs.

\begin{figure}[htbp]
\centering
\includegraphics[width=\linewidth]{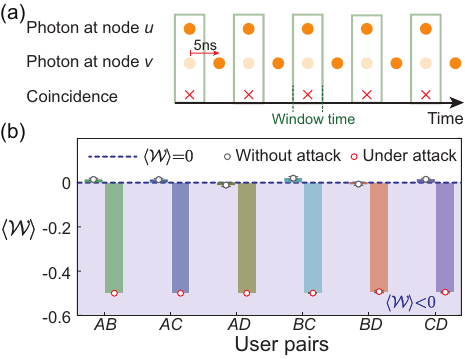}
\caption{\textbf{Time-shift attack on entanglement witness measurement. (a)}~Schematic illustration of the time-shift attack. In the absence of manipulation, photons at nodes $u$ and $v$ produce valid coincidence events within the selected time window. Under attack, a temporal delay of $\sim$5~ns is introduced on the photon at node $v$, while the coincidence window remains fixed. As a result, coincidence events fall outside the window and are not registered, leading to biased correlation estimates. \textbf{(b)}~Measured entanglement witness values $\langle \mathcal{W} \rangle$ for the separable state $\ket{e_ue_v}$ for all six user pairs. The gray and red circles denote the results obtained without and under the time-shift attack, respectively.}
\label{fig:ew_attack}
\end{figure}

\begin{figure*}[ht!bp]
\centering
\includegraphics[width=\linewidth]{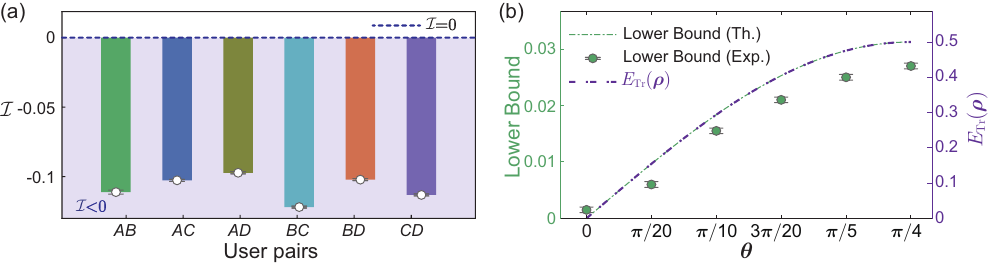}
\caption{\textbf{Results of MDI entanglement Verification and quantification for the time-bin entangled states $\ket{\Phi(\theta)}$. (a)~} The measured MDI witness values $\mathcal I(\rho_{u, v})$ for all six links. \textbf{(b)~}MDI entanglement quantification for tunable time-bin states $\ket{\Phi(\theta)}$ in link $(A, B)$. The green dashed curve shows the theoretical lower bound, while the purple dashed curve shows the corresponding trace-distance entanglement measure $E_{\mathrm{Tr}}(\rho)$, plotted on the right vertical axis. Green circles denote the experimentally measured lower bounds.}
\label{Fig:MDI}
\end{figure*}

While these results confirm high-quality entanglement across all links, such certification relies on well-characterized measurement devices~(green shaded region in Fig.~\ref{fig:setup}~(c)). To illustrate this limitation, we experimentally demonstrate a time-shift attack in the entanglement witness measurement using the separable state $\ket{e_ue_v}$ as a test case. In the absence of the attack, the witness values obtained without the attack are close to zero for all user pairs, as expected for a separable state, as shown by the gray circles in Fig.~\ref{fig:ew_attack}~(b). Small positive and negative deviations are attributed to experimental imperfections in state preparation and measurement. The attack mechanism is illustrated in Fig.~\ref{fig:ew_attack}~(a). For the verification between nodes $u$ and $v$, the attack is implemented by shifting the arrival time (by $\sim$5~ns) of the photon at node $v$ in selected measurement bases, while keeping the coincidence window fixed. Consequently, coincidence events in these bases fall outside the detection window and are not registered. For example, in the measurement of $\sigma_x \otimes \sigma_x$, the time shift is applied to the bases $\ket{+}\ket{-}$ and $\ket{-}\ket{+}$, where $\ket{\pm}=(\ket{H}\pm\ket{V})/\sqrt{2}$ denote the eigenstates of $\sigma_x$. This suppresses their coincidence counts and yields an apparent expectation value $\langle \sigma_x \otimes \sigma_x \rangle = 1$, whereas the correct value for the separable state $\ket{e_u e_v}$ is 0. Similar attacks lead to $\langle\sigma_y \otimes \sigma_y\rangle=-1$ and $\langle\sigma_z \otimes \sigma_z\rangle=1$. As shown by the red circles in Fig.~\ref{fig:ew_attack}~(b), the time-shift attack drives the estimated witness values to $\langle \mathcal{W} \rangle = -0.5$, far below zero and well beyond the statistical uncertainty of the no-attack results, thereby falsely certifying the separable state $\ket{e_ue_v}$ as entangled. More details are provided in the Supplementary Materials.

To overcome this limitation, we implement the MDI-EW, which removes any assumptions about the measurement apparatus. In the MDI framework, $\mathcal W$ in Eq.~\ref{Eq:EW} can be written in the~(nonunique) form of $\mathcal W=\sum_{s, t}\beta_{s, t}\tau_s^\top\otimes\omega_t^\top$, where $\beta_{s,t}$ are real coefficients and $\tau_s$, $\omega_t$ are auxiliary input states acting on the Hilbert spaces $\mathcal H_u$ and $\mathcal H_v$, respectively. For a given user pair $(u,v)$ sharing the state $\rho_{uv}$, the inputs $\tau_s$ and $\omega_t$ are jointly projected with $\rho_{uv}$ onto the maximally entangled state $\ket{\Phi^+}$. The corresponding success probability $P(1,1|\tau_s,\omega_t)$ is used to construct the MDI witness
\begin{equation}
\mathcal I=\sum_{s, t}\beta_{s, t}P(1, 1|\tau_s, \omega_t),
\end{equation}    
which satisfies $\mathcal I\geq0$ for all separable states $\sigma_{uv}$ and $\mathcal I<0$ for entangled states $\rho_{uv}$.

In our implementation, the trusted input states are encoded in the polarization DoF of the same photons, eliminating the need for additional ancillary photons or optical modes. As shown in Fig.~\ref{fig:setup}~(d), a combination of a HWP and a QWP is used to prepare the sets of input states $\{\tau_s(\omega_t)\}=\{\ket{\psi}\bra{\psi}\}$ with $\ket{\psi}\in\{\ket{H}, \ket{V}, \ket{+}, \ket{-}, \ket{L}, \ket{R}\}$, where $\ket{L(R)}=(\ket{H}\pm i\ket{V})/\sqrt{2}$. These inputs realize the decomposition of the witness operator in Eq.~\ref{Eq:EW}, allowing the witness value to be reconstructed from the measured probabilities $P(1,1|\tau_s,\omega_t)$.

The Bell-state measurement~(BSM) is implemented by projecting onto $\ket{\Phi^+}=\frac{1}{\sqrt{2}}(\ket{He}+\ket{Vl})$, corresponding to a hybrid entangled state between the polarization and time-bin DoF. This is achieved by setting HWP$_1=0\degree$ and HWP$_2=22.5\degree$. Notably, the set of optical elements that must be well characterized in the MDI measurement~(green shaded region in Fig.~\ref{fig:setup}~(d)) is significantly reduced compared to that required for conventional projective measurements~(green shaded region in Fig.~\ref{fig:setup}~(c)). As shown in Fig.~\ref{Fig:MDI}~(a), the measured MDI witness values $\mathcal I(\rho_{u, v})$ for all six links are negative, lying below the separable bound and confirming entanglement across the entire network. Specifically, we obtain $\mathcal{I} = -0.111 \pm 0.001$, $-0.103 \pm 0.001$, $-0.097 \pm 0.001$, $-0.122 \pm 0.001$, $-0.102 \pm 0.001$, and $-0.113 \pm 0.001$ for the links $AB$, $AC$, $AD$, $BC$, $BD$, and $CD$, respectively, corresponding to violations of the separable bound by more than 100 standard deviations. More details regarding the experimental realization of MDI measurements are provided in the Supplementary Materials.

The MDI framework further enables direct quantification of entanglement. 
We consider the trace-distance entanglement measure, defined as the minimal distance between the state $\rho$ and the set of separable states $\Omega$~\cite{vedral1997PRL},
\begin{equation}
E_{\mathrm{Tr}}(\rho)=\min_{\sigma\in\Omega}\frac{1}{2}\mathrm{Tr}|\sigma-\rho|.
\end{equation}
Although $E_{\mathrm{Tr}}(\rho)$ is not directly accessible in the MDI setting, it admits a measurable lower bound that can be evaluated from the same experimental statistics used for the witness. Following Ref.~\cite{sun2024PRL}, we obtain
\begin{equation}
E_{\mathrm{Tr}}(\rho)\ge -\frac{\mathcal{I}(\rho)}{2\mathrm{Tr}|\mathcal W^\top|}.
\end{equation}
This bound depends only on the experimentally measured MDI witness value $\mathcal{I}(\rho)$, enabling entanglement quantification to be performed directly from the same dataset used for entanglement certification, without requiring additional measurements. Further details are provided in the Supplementary Materials.

Experimentally, we apply this protocol to a family of time-bin entangled states shared between user $A$ and $B$
\begin{equation}
\ket{\Phi(\theta)}=\cos\theta\ket{e_A e_B}+\sin\theta\ket{l_A l_B},
\quad 0\le\theta\le\frac{\pi}{4},
\label{eq:tunable_timebin_state}
\end{equation}
which interpolate from the separable state $\ket{e_A e_B}$ to the maximally entangled state $(\ket{e_A e_B}+\ket{l_A l_B})/\sqrt{2}$ as $\theta$ increases. In the experiment, we prepare six states with $\theta= 0, \pi/20, \pi/10, 3\pi/20, \pi/5$ and $\pi/4$. As shown in Fig.~\ref{Fig:MDI}~(b), the experimental measured lower bound~(green dots) exhibit a clear monotonic increase with the parameter $\theta$, in good agreement with the theoretical prediction~(green dashed line). Although the MDI lower bound lies below the actual entanglement measure $E_\text{Tr}(\rho)$~(purple dashed line), as expected, it faithfully follows the same trend, demonstrating that the protocol captures the variation in entanglement strength.

In conclusion, we experimentally demonstrate MDI entanglement verification and quantification in a time-bin FCQN. Using a broadband PPLNOI source and wavelength multiplexing, we distribute six entangled links among four users over 20-km fiber channels and verify their robustness without active stabilization of the long-distance links. By encoding trusted inputs in the polarization DoF of the same photons, we realize an MDI protocol without ancillary photons or additional optical resources. Within this framework, both entanglement verification and quantification are obtained from the same measurement dataset, providing a resource-efficient approach to reliable entanglement characterization. Our implementation is inherently scalable. The PPLNOI platform can be engineered to generate broadband photon pairs~\cite{javid2021PRL,fang2026OEO}, enabling wavelength multiplexing to distribute many entangled links from a single integrated source, while the MDI scheme eliminates the need for ancillary photons and relaxes requirements on device characterization, making it well suited for large-scale quantum networks. The MDI verification and quantification framework can be naturally extended to multipartite scenarios~\cite{Zhao2016PRA,Shahandeh2017PRL}, and is compatible with recent experimental progress in time-bin multiphoton entanglement~\cite{Reimer2016Science,Lo2023QST,fang2026arXiv}, opening a route toward scalable and complex quantum communication tasks. 



\clearpage

\onecolumngrid
\setcounter{equation}{0}
\setcounter{figure}{0}
\renewcommand{\thefigure}{S\arabic{figure}}
\renewcommand{\theequation}{S\arabic{equation}}


\begin{center}
\Large\bfseries
Supplementary Materials for ``Measurement-Device-Independent Entanglement Quantification in a Fully Connected Time-Bin Quantum Network''
\end{center}

\section{Design, fabrication and characterization of the periodically poled lithium niobate on insulator~(PPLNOI) waveguide} \label{sup:ppln}
\subsection{Phase-matching design of the PPLNOI waveguide}

\begin{figure}[htbp]
\centering
\includegraphics[width=0.6\linewidth]{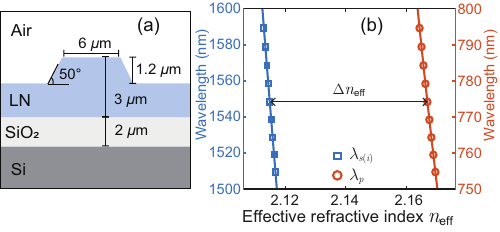}
\caption{\textbf{(a)}~Designed cross section of the PPLNOI ridge waveguide . 
\textbf{(b)}~Simulated effective refractive index of the designed waveguide.
}
\label{fig:wg}
\end{figure}

We consider spontaneous parametric down-conversion (SPDC) in a straight ridge waveguide fabricated on a lithium-niobate-on-insulator (LNOI) platform, consisting of a 3-$\mu$m-thick $z$-cut 5\% MgO-doped lithium niobate (LN) thin film bonded to a 0.5-mm-thick silicon substrate via a 2-$\mu$m-thick SiO$_2$ layer. The waveguide cross section is shown in Fig.~\ref{fig:wg}~(a), with a width of $\sim6~\mu$m, an etch depth of $\sim1.2~\mu$m, and a sidewall angle of $\sim50^\circ$. This geometry supports fundamental transverse-magnetic (TM$_{00}$) modes at both the pump and down-converted wavelengths, enabling type-0 SPDC. 

In the degenerate SPDC process, a pump photon ($p$) at wavelength $\lambda_p=775$~nm is converted into two photons, i.e., signal ($s$) and idler ($i$), with equal wavelengths $\lambda_s=\lambda_i\approx1550$~nm, satisfying the energy conservation condition $1/\lambda_p=1/\lambda_s+1/\lambda_i$. For this geometry, the guided modes are calculated numerically using a finite-element method, from which the effective refractive indices $n_{\mathrm{eff}}(\lambda)$ are obtained. These determine the propagation constants $\beta_{\mu} = 2\pi n_{\mathrm{eff}}(\lambda_{\mu})/\lambda_{\mu}$ for $\mu=p,s,i$.

The phase-matching condition for the TM$_{00}\rightarrow$TM$_{00}$+TM$_{00}$ process is given by the phase mismatch
$\Delta\beta=\beta_s+\beta_i-\beta_p-\beta_{\Lambda}$,
where $\beta_{\Lambda}=2\pi/\Lambda$ is the reciprocal lattice vector introduced by periodic poling with period $\Lambda$. Efficient nonlinear interaction requires $\Delta\beta\approx0$. Using the numerically calculated effective refractive indices at $\lambda_p=775$~nm and $\lambda_{s(i)}=1550$~nm, the effective index mismatch between the corresponding modes is $\Delta n_{\mathrm{eff}}=0.0516$, and the required poling period is determined to be $\Lambda=\frac{0.775 \mu \text{m}}{\Delta n_{\mathrm{eff}}}\approx15~\mu$m.

\subsection{Fabrication of the PPLNOI waveguide}
\begin{figure}[htbp]
\centering
\includegraphics[width=0.6\linewidth]{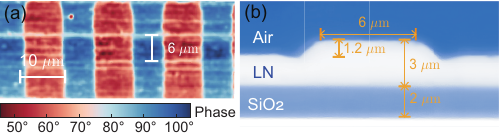}
\caption{Fabrication. 
\textbf{(a)}~PFM image of the periodically poled region, confirming clear domain inversion with a duty cycle of about $45\pm2\%$.
\textbf{(b)}~Microscope image of the fabricated waveguide end facet. 
}
\label{fig:pfm}
\end{figure}

The fabrication process begins with periodic poling of the LN thin film to achieve quasi-phase matching. The electrode pattern for domain inversion is defined by laser direct writing, followed by metal deposition and lift-off to form the poling electrodes. High-voltage pulses are then applied to induce ferroelectric domain inversion in the designated regions. The resulting periodically poled structure is characterized by piezoelectric force microscopy (PFM), as shown in Fig.~\ref{fig:pfm}~(a), confirming uniform domain inversion with a duty cycle of $45\pm2\%$. The ridge waveguide is subsequently fabricated within the poled region. A $10~\mu$m-thick photoresist (AR-4400) is spin-coated as the etching mask, and the waveguide pattern is defined by laser direct writing. The pattern is then transferred into the lithium niobate thin film via inductively coupled plasma reactive-ion etching (ICP-RIE), forming the ridge-waveguide structure. The cross section of the fabricated waveguide is shown in Fig.~\ref{fig:pfm}~(b), which agrees well with the designed geometry and confirms the realization of the intended ridge profile.

\subsection{Classical and quantum characterization of the phase matching}
\begin{figure}[htbp]
\centering
\includegraphics[width=1\columnwidth]{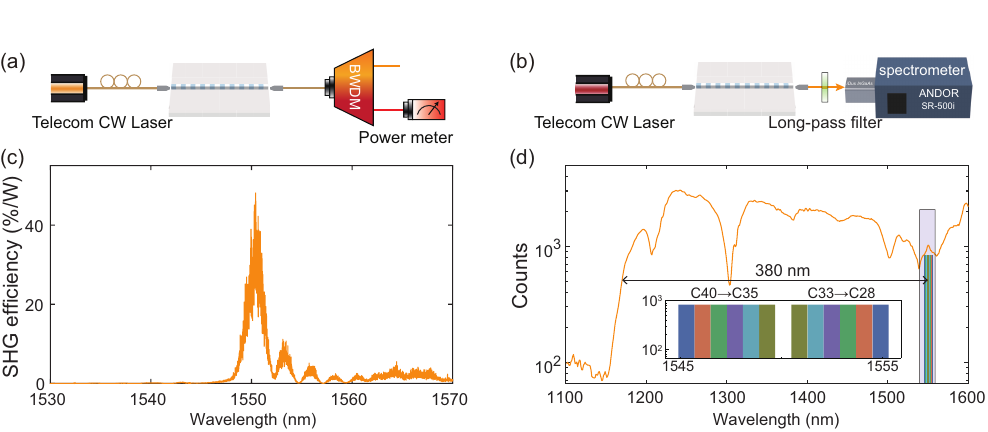}
\caption{\textbf{(a)}~Experimental setup for second-harmonic generation~(SHG) measurement, where a telecom continuous-wave laser is coupled into the waveguide and the generated SH signal is separated using a band wavelength-division multiplexer~(BWDM) and measured with a power meter. \textbf{(b)}~Experimental setup for the spontaneous parametric down-conversion~(SPDC) spectrum characterization, where the photon pair is filtered by a long-pass filter and analyzed using a spectrometer. \textbf{(c)}~Measured SHG efficiency as a function of the wavelength of first-harmonic~(FH) light, showing a phase-matching peak around 1550~nm. \textbf{(d)}~Measured SPDC spectrum under a 775~nm pump, exhibiting a broad bandwidth spanning the telecom regime. The inset highlights the selected DWDM channels (C40–C35 and C33–C28), which form six frequency-correlated photon-pair channels used for wavelength-multiplexed entanglement distribution.}
\label{fig:shg}
\end{figure}

To verify the phase-matching condition, we first perform second-harmonic generation (SHG) measurements by scanning the wavelength of the first-harmonic (FH) light in the telecom band. As shown in Fig.~\ref{fig:shg}~(a), the FH light, with input power $P_\text{FH}$, is provided by a continuous-wave tunable laser~(Santec TSL-550). A polarization controller (PC) is used to prepare the FH light in the TM$_{00}$ mode, matching the waveguide mode. The light is coupled into the PPLNOI waveguide via a lensed fiber (LF), and the generated second-harmonic (SH) signal, together with residual FH light, is collected at the output using another LF. The fiber-to-waveguide coupling efficiency is approximately 3~dB per facet. To isolate the SH light, a 775/1550~nm band wavelength-division multiplexer~(BWDM) is employed to efficiently suppress the residual FH light. The SH power $P_\text{SH}$ is then measured using a power meter. The normalized SHG efficiency is calculated by
\begin{equation}
\eta_{\mathrm{SHG}} = \frac{P_{\mathrm{SH}} / \eta_{\mathrm{SH}}}{\left(P_{\mathrm{FH}} / \eta_{\mathrm{FH}}\right)^2},
\end{equation}
where $\eta_{\mathrm{FH}}$ and $\eta_{\mathrm{SH}}$ denote the transmission efficiencies for the FH and SH light, respectively. For our PPLNOI device, $\eta_{\mathrm{FH}}=55\%$ and $\eta_{\mathrm{SH}}=50\%$. The chip is mounted on a thermoelectric cooler stabilized at $26\degree\mathrm{C}$, and the SHG spectrum is obtained by sweeping the FH wavelength. As shown in Fig.~\ref{fig:shg}~(c), the SHG phase-matching peak is observed at $\sim1550$~nm, with a normalized conversion efficiency of $48.1\%/\mathrm{W}$. This confirms that the waveguide is phase matched for degenerate SPDC at telecom wavelengths under a $775~\mathrm{nm}$ pump.

We further characterize the SPDC spectrum of the PPLNOI waveguide under a 775~nm pump. The experimental setup is shown in Fig.~\ref{fig:shg}~(b). The photons generated via SPDC are collected by a LF and passed through a long-pass filter to suppress residual pump light. The spectrum is then measured using a high-sensitivity spectrometer (Andor SR-500i). Detailed experimental descriptions of the setup can be found in our previous work~\cite{fang2026OEO}. The measured SPDC spectrum, shown in Fig.~\ref{fig:shg}~(d), exhibits a broad bandwidth centered in the telecom regime. In particular, the spectrum extends over approximately 1126~nm~(corresponding to 126~THz), reflecting the wide phase-matching window of the PPLNOI waveguide. The broadband emission enables simultaneous access to multiple frequency-correlated channel pairs, providing the basis for wavelength-multiplexed distribution of entanglement in the quantum network.

 
\subsection{Characterization of the wavelength-multiplexed photon-pair source}
\label{sec:selectedDWDM}
To characterize the performance of the broadband SPDC source across different wavelength channels, we pump the PPLNOI waveguide using the same pulsed laser as in the main experiment. A $100~\mathrm{GHz}$ dense wavelength-division multiplexing (DWDM) system is used to separate the broadband emission and select frequency-correlated channel pairs. Owing to energy conservation, signal and idler photons are generated symmetrically about the degenerate wavelength. The DWDM system therefore defines correlated channel pairs $(s_j,i_j)$, where $j=1,2,\dots,N$ labels the channel index with increasing spectral separation. In the experiment, six symmetric DWDM channel pairs are selected around the degeneracy, with signal channels C35–C40 paired with idler channels C33–C28, respectively. Here, CXX denotes the standard ITU-T DWDM channel index in the C-band on a $100~\mathrm{GHz}$ grid. These channel pairs are used for both source characterization and the subsequent network implementation.

\begin{figure}[htbp]
\centering
\includegraphics[width=\linewidth]{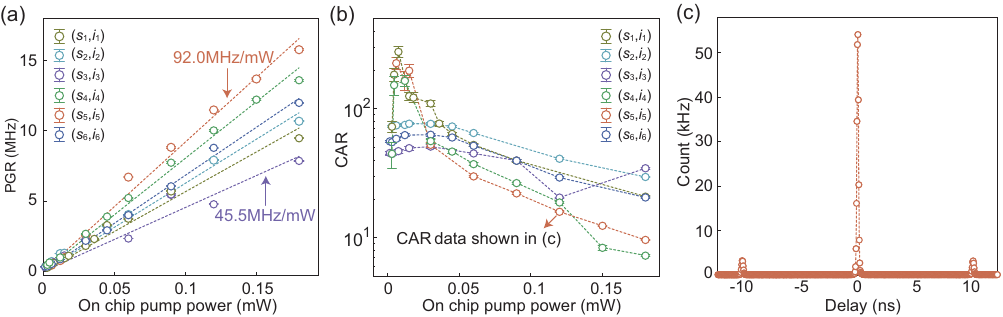}
\caption{\textbf{Performance characterization of the wavelength-multiplexed SPDC source. (a)}~Pair generation rate (PGR) as a function of on-chip pump power for six channel pairs $(s_j,i_j)$. The data exhibit linear scaling, with fitted slopes ranging from 45.5 to 92.0~MHz/mW. \textbf{(b)}~Coincidence-to-accidental ratio (CAR) as a function of pump power for the same channel pairs. \textbf{(c)}~Representative coincidence histogram for channel pair $(s_5,i_5)$, showing the coincidence peak at zero delay and side peaks corresponding to accidental events, from which the CAR is extracted.}
\label{fig:PGR}
\end{figure}

The source performance is evaluated in terms of the pair generation rate~(PGR) and the coincidence-to-accidental ratio~(CAR). For each channel pair $(s_j, i_j)$, the signal ($N_{s_j}$), idler ($N_{i_j}$), and coincidence ($N_{c_j}$) count rates are recorded within a coincidence window of 1~ns. The PGR$_j$ is defined as  
\begin{equation}
    \text{PGR}_j = \frac{N_{s_j}N_{i_j}}{N_{c_j}}
\end{equation}
As shown in Fig.~\ref{fig:PGR}~(a), the PGR$_j$ exhibits a linear dependence on the on-chip pump power for all channel pairs. From linear fits, the extracted slopes $k_{j}$ range from 45.5~MHz/mW to 92.0~MHz/mW. We observe noticeable variations in PGR$_j$ across different channels. This is mainly attributed to non-ideal DWDM filtering, including imperfect spectral symmetry between signal and idler channels and finite channel isolation. 

The CAR for each channel pair is defined as
\begin{equation} 
\text{CAR}_j = \frac{C_{\max,j} - C_{\text{acc},j}}{C_{\text{acc},j}},
\end{equation} 
where $C_{\max,j}$ denotes the coincidence counts at zero delay and $C_{\text{acc},j}$ is the average accidental count. Fig.~\ref{fig:PGR}~(c) shows a representative coincidence histogram illustrating the procedure used to extract $\text{CAR}_j$ measured for the channel pair $(s_5,i_5)$ at a pump power of $\sim 0.12~\mathrm{mW}$. The peak at zero delay corresponds to true photon-pair coincidences, while the side peaks at $\pm 10~\mathrm{ns}$ (set by the laser repetition period) represent accidental coincidences and are used to estimate $C_{\text{acc},j}$. The CAR values for all data points in Fig.~\ref{fig:PGR}~(b) are obtained using this procedure.

The CAR for each channel pair, $\text{CAR}_j$, as a function of on-chip pump power is shown in Fig.~\ref{fig:PGR}~(b). For all channel pairs, $\text{CAR}_j$ decreases with increasing pump power due to the growing contribution of accidental coincidences and multi-pair emissions. At a pump power of approximately $0.1~\mathrm{mW}$, all channel pairs exhibit $\text{CAR}_j > 20$, which indicates a sufficiently high signal-to-noise ratio for measurements.

\section{Construction of the four-user time-bin fully connected quantum network~(FCQN)}
\subsection{Generation of time-bin entangled photons}

\begin{figure}[htbp]
\centering
\includegraphics[width=0.5\linewidth]{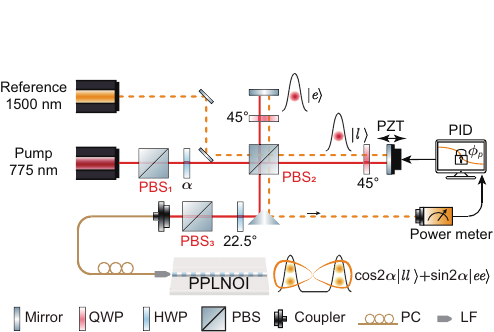}
\caption{Experimental setup to generate time-bin entangled photons.}
\label{fig:timebin_generate}
\end{figure}

As shown in Fig.~\ref{fig:timebin_generate}, a dual-pulse pump is prepared using an unbalanced Mach–Zehnder interferometer~(UMZI). A pulsed laser first passes through PBS$_1$ and a half-wave plate~(HWP) set as $\alpha = 22.5^\circ$, preparing the polarization state $(\ket{H}+\ket{V})/\sqrt{2}$. The two polarization components are then separated by PBS$_2$ into the short and long arms of the interferometer, corresponding to the early ($\ket{e_p}$) and late ($\ket{l_p}$) time bins, respectively. After propagating through the interferometer, both components pass twice through a quarter-wave plate~(QWP) set at $45^\circ$, which flips their polarizations and enables recombination at PBS$_2$. This process results in a coherent superposition of the two time bins, with a relative phase $\phi_p$ determined by the path-length difference of the UMZI. After passing through an additional HWP at $22.5^\circ$ and PBS$_3$, the pump polarization is projected onto $\ket{H}$, yielding the dual-pulse state $\frac{1}{\sqrt{2}}(\ket{e_p}+e^{i\phi_p}\ket{l_p})\otimes\ket{V}$. A 1500~nm reference beam is injected into the UMZI, and the phase is actively stabilized at $\phi_p = 0$ via a piezoelectric transducer~(PZT) to prepare the quantum state $\frac{1}{\sqrt{2}}(\ket{e_p}+\ket{l_p})\otimes\ket{V}$. The dual-pulse pump is then injected into the PPLNOI waveguide after polarization control. The polarization controller rotates the pump polarization to match the type-0 SPDC. Through SPDC, each pump pulse probabilistically generates a photon pair in the corresponding temporal mode, yielding the time-bin entangled state $(\ket{e_s e_i}+e^{i\phi_p}\ket{l_s l_i})/\sqrt{2}$. Owing to the broadband phase-matching of the PPLNOI waveguide, this process simultaneously generates time-bin entangled photon pairs across multiple frequency-correlated channels, providing the entanglement resources for the subsequent wavelength-multiplexed FCQN.

More generally, biased entangled time-bin states 
\begin{equation}
    \ket{\Phi(\theta)}=\cos\theta\ket{e_s e_i}+\sin\theta\ket{l_s l_i},
\quad 0\le\theta\le\frac{\pi}{4}
\end{equation}
can be prepared by rotating the HWP placed before the pump light enters the UMZI. In this case, the polarization state incident on PBS$_2$ is no longer an equal superposition of $\ket{H}$ and $\ket{V}$, resulting in unequal optical amplitudes in the short and long arms. Consequently, the balanced dual-pulse pump is generalized to an unbalanced superposition of early and late time bins. The parameter $\theta$ is related to the HWP rotation angle $\alpha$ by $\theta = 2\alpha$. The six values used in the experiment are
$\theta=0,\ \pi/20,\ \pi/10,\ 3\pi/20,\ \pi/5,\ \pi/4$, corresponding to HWP angle $\alpha=0^\circ, 4.5^\circ, 9.0^\circ, 13.5^\circ, 18.0^\circ, 22.5^\circ$.

\subsection{Wavelength-multiplexed distribution of time-bin entangled photons}
\begin{figure}[htbp]
\centering
\includegraphics[width=0.5\linewidth]{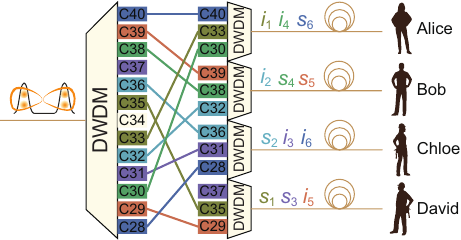}
\caption{\textbf{Wavelength-multiplexed distribution of time-bin entangled photon pairs.} The broadband SPDC photon source is demultiplexed into multiple DWDM channels, where frequency-correlated signal–idler pairs $(s_j,i_j)$ are symmetrically located about the degenerate wavelength. The channels are selectively routed and recombined using four DWDMs to distribute photons to four users (Alice, Bob, Chloe, and David). Each user receives three wavelength channels via a single fiber link, and each channel pair $(s_j,i_j)$ is assigned to a distinct user pair, thereby establishing six entangled links and realizing a FCQN.}
\label{Fig:Demux}
\end{figure}

The broadband time-bin entangled photon pairs generated from the PPLNOI waveguide are demultiplexed using the same DWDM described in Sec.~\ref{sec:selectedDWDM}. After demultiplexing, the signal and idler channels are selectively routed and recombined using DWDM multiplexers to distribute the entangled photons to different users over 20-km fiber links. In our experiment, four DWDMs are used to multiplex the assigned wavelength channels for each user into a single fiber link, as illustrated in Fig.~\ref{Fig:Demux}. Specifically, Alice receives ${i_1, i_4, s_6}$, Bob receives ${i_2, s_4, s_5}$, Chloe receives ${s_2, i_3, i_6}$, and David receives ${s_1, s_3, i_5}$, such that each user receives three wavelength channels. With this channel allocation, each correlated pair $(s_j,i_j)$ is distributed to a distinct pair of users, thereby establishing one entangled link $(u,v)$. The six channel pairs are thus mapped to the user pairs $(A,B)$, $(A,C)$, $(A,D)$, $(B,C)$, $(B,D)$, and $(C,D)$, respectively, realizing a FCQN.

\section{Characterization of time-bin entanglement with projective measurement}

\subsection{Experimental setups for projective measurements}

\begin{figure}[htbp]
\centering
\includegraphics[width=0.5\linewidth]{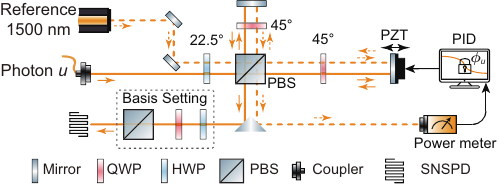}
\caption{Experimental setup to perform projective measurement on time-bin qubit.}
\label{fig:projective_setup}
\end{figure}
\begin{figure*}[htbp]
\centering
\includegraphics[width=0.8\linewidth]{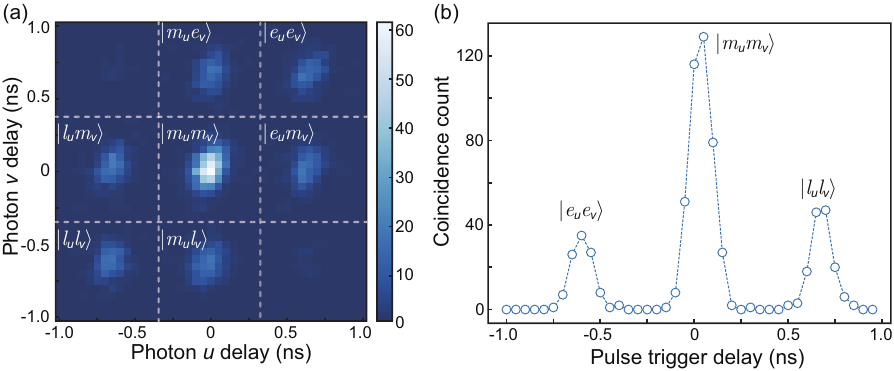}
\caption{\textbf{(a)}~Two-dimensional coincidence map obtained by scanning the relative delays of photons $u$ and $v$, where distinct peaks correspond to different time-bin basis states. \textbf{(b)}~Coincidence counts as a function of detection time relative to the pump pulse trigger, showing three well-resolved peaks corresponding to $\ket{e_u e_v}$, $\ket{m_u m_v}$, and $\ket{l_u l_v}$. The central peak $\ket{m_u m_v}$ arises from interference between the early–long and late–short paths. The clear temporal separation of these peaks enables selective postselection of the middle time-bin component for time-bin-to-polarization conversion.}
\label{fig:delay_scan}
\end{figure*}

We perform projective measurements of time-bin entanglement using the setup shown in Fig.~\ref{fig:projective_setup}. The key idea is to coherently map the time-bin degree of freedom onto the polarization degree of freedom, thereby enabling standard polarization analysis. To illustrate the operation of the setup, we present a step-by-step transformation in Eq.~\ref{eq:DOFConversion}, taking the time-bin Bell state 
$\ket{\Phi^+_\text{time}} = \frac{1}{\sqrt{2}} \left( \ket{e_u e_v} + \ket{l_u l_v} \right)$ as an example. The polarization of both photons is initially rotated to $\ket{H}$ by polarization controller.
\begin{equation}
\begin{aligned}
\ket{\Phi^+_\text{time}}&\xrightarrow{\mathrm{HWP}@22.5^\circ}  
\frac{1}{2\sqrt{2}} \left( \ket{e_u e_v} + \ket{l_u l_v} \right) 
\otimes \left( \ket{H_u H_v} + \ket{H_u V_v} + \ket{V_u H_v} + \ket{V_u V_v} \right) \\
& \xrightarrow{\mathrm{UMZI}} 
\frac{1}{2\sqrt{2}} \left( e^{i (\phi_u + \phi_v)}\ket{m_u m_v} + e^{i ( \phi_u + \phi_v)} \ket{l_u l_v} \right) \otimes \ket{V_u V_v} \\
& \qquad + \frac{1}{2\sqrt{2}} \left( e^{i\phi_u}\ket{m_u e_v} + e^{i \phi_u}\ket{l_u m_v} \right) \otimes \ket{V_u H_v} \\ 
& \qquad + \frac{1}{2\sqrt{2}} \left( e^{i\phi_v}\ket{e_u m_v} + e^{i \phi_v}\ket{m_u l_v} \right) \otimes \ket{H_u V_v} \\
& \qquad + \frac{1}{2\sqrt{2}}\left( \ket{e_u e_v} + \ket{m_u m_v} \right) \otimes \ket{H_u H_v} \\
& \xrightarrow[\phi_u+\phi_v=0]{\text{Postselection}} 
\frac{1}{\sqrt{2}} \left( \ket{H_u H_v} + \ket{V_u V_v} \right)\otimes\ket{m_u m_v}.
\end{aligned}
\label{eq:DOFConversion}
\end{equation}
According to Eq.~\ref{eq:DOFConversion}, after passing through the UMZI, the biphoton state evolves into a superposition of multiple time-bin basis states. To experimentally resolve these components, we perform coincidence measurements by scanning the relative delays of photons $u$ and $v$. The resulting two-dimensional coincidence map is shown in Fig.~\ref{fig:delay_scan}~(a), where distinct peaks correspond to different time-bin basis states. 

Fig.~\ref{fig:delay_scan}~(b) shows the coincidence counts as a function of the detection time relative to the pump pulse trigger. Three well-resolved peaks are observed, corresponding to the early–early $\ket{e_u e_v}$, middle–middle $\ket{m_u m_v}$, and late–late $\ket{l_u l_v}$ time-bin components. The central peak $\ket{m_u m_v}$ arises from the interference between the early–long and late–short paths. The clear temporal separation of these peaks confirms that the $\ket{m_u m_v}$ component can be selectively postselected in the experiment, which is essential for implementing the time-bin to polarization conversion.

\subsection{Quantum tomography of time-bin entangled photons}
\begin{figure*}[h!t]
\centering
\includegraphics[width=0.85\textwidth]{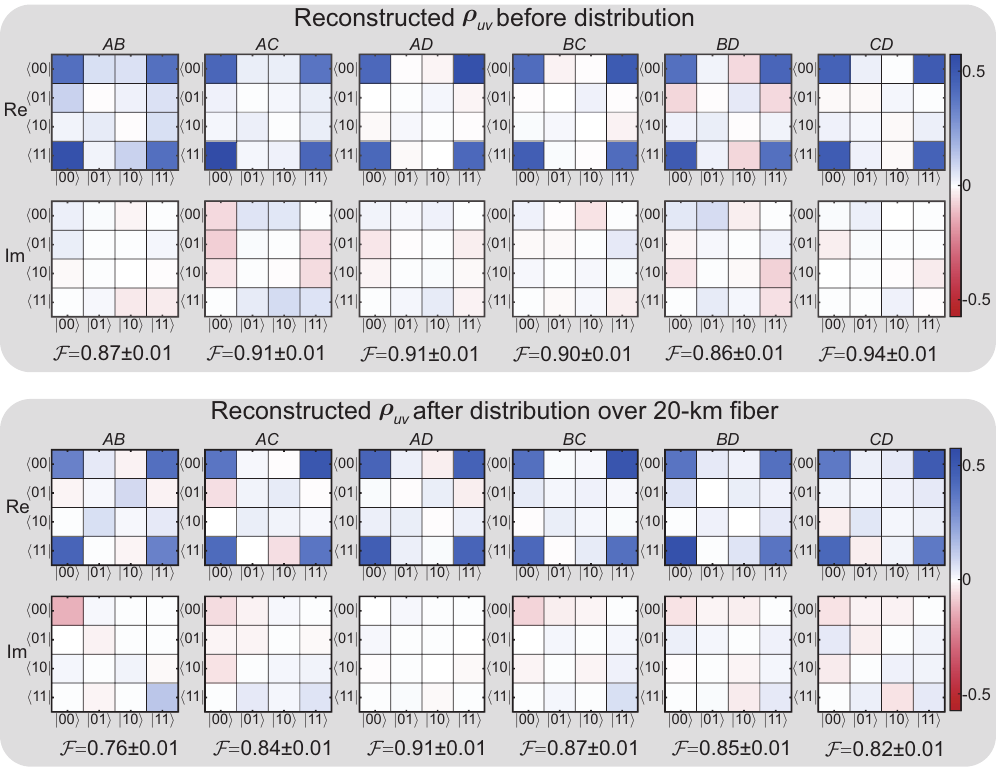}
\caption{\textbf{Reconstructed two-photon density matrices for all user pairs before and after distribution.} Real~(top row) and imaginary~(bottom row) parts of the reconstructed density matrices $\rho_{uv}$ for the six user pairs ($AB$, $AC$, $AD$, $BC$, $BD$, and $CD$). The upper panel shows the results before distribution, while the lower panel shows the results after transmission over 20-km fiber links. The fidelities with respect to the ideal state $\ket{\Phi^+}$ are indicated below each matrix.}
\label{fig:tomo}
\end{figure*}

With the projective-measurement scheme described above, the time-bin qubits are coherently mapped onto the polarization DoF and analyzed via standard two-qubit quantum state tomography. The measurements are performed by projecting each photon onto a set of polarization bases $\{\ket{H}, \ket{V}, \ket{+}, \ket{-}, \ket{L}, \ket{R}\}$, where $\ket{\pm} = (\ket{H} \pm \ket{V})/\sqrt{2}$ and $\ket{L(R)} = (\ket{H} \pm i\ket{V})/\sqrt{2}$. Coincidence counts are recorded for all combinations of local measurement settings, yielding an overcomplete set of measurement outcomes. The two-photon density matrix $\rho_{uv}$ is then reconstructed using a maximum-likelihood estimation method, ensuring a physical~(positive semidefinite) density matrix with unit trace. The real and imaginary parts of the reconstructed density matrices for all user pairs are shown in Fig.~\ref{fig:tomo}, both before and after distribution over 20-km fiber links. The corresponding state fidelities with respect to the ideal Bell state $\ket{\Phi^+}$ are indicated below each panel. High fidelities are observed for all six links before transmission, and remain consistently high after transmission, demonstrating the robust preservation of time-bin entanglement over long-distance fiber distribution in the network.

\section{Entanglement witness and time-shift attack}\label{sup:attack}
For the target Bell state $\ket{\Phi^+_\text{pol}} = (\ket{H_uH_v} + \ket{V_uV_v})/\sqrt{2}$, the conventional entanglement witness used in this work is defined as
\begin{equation}
    \mathcal W = \frac{\mathbb{I}_4}{2} - \ket{\Phi^+}\bra{\Phi^+}.
\end{equation}
Using the Pauli operators $\sigma_x$, $\sigma_y$, and $\sigma_z$, the projector onto $\ket{\Phi^+}$ can be expressed as
\begin{equation}
    \ket{\Phi^+}\bra{\Phi^+}
    = \frac{1}{4}\left(
    \mathbb I_2 \otimes \mathbb I_2
    + \sigma_x \otimes \sigma_x
    - \sigma_y \otimes \sigma_y
    + \sigma_z \otimes \sigma_z
    \right),
\end{equation}
which yields
\begin{equation}\label{Eq:EWsupp}
    \mathcal W = \frac{1}{4}\left(
    \mathbb I_2 \otimes \mathbb I_2
    - \sigma_x \otimes \sigma_x
    + \sigma_y \otimes \sigma_y
    - \sigma_z \otimes \sigma_z
    \right).
\end{equation}
The expectation value of the witness can therefore be obtained from local measurements as
\begin{equation}
    \langle\mathcal W\rangle=\mathrm{Tr}[\mathcal W\rho_{uv}]
    = \frac{1}{4}\left(
    1 - \langle \sigma_x \sigma_x \rangle
    + \langle \sigma_y \sigma_y \rangle
    - \langle \sigma_z \sigma_z \rangle
    \right).
\end{equation}
For any separable state $\sigma_{uv}$, one has $\langle\mathcal W\rangle\ge 0$, while a negative value certifies entanglement. The correlation terms $\sigma_i \sigma_i$ ($i = x,y,z$) can be decomposed in terms of the projectors onto the eigenstates of the corresponding Pauli operators, i.e.,
\begin{equation}
\begin{split}
\sigma_x \otimes \sigma_x =& \ket{++}\bra{++} - \ket{+-}\bra{+-} - \ket{-+}\bra{-+} + \ket{--}\bra{--}, \\
\sigma_y \otimes \sigma_y =& \ket{LL}\bra{LL} - \ket{LR}\bra{LR} - \ket{RL}\bra{RL} + \ket{RR}\bra{RR}, \\
\sigma_z \otimes \sigma_z =& \ket{HH}\bra{HH} - \ket{HV}\bra{HV} - \ket{VH}\bra{VH} + \ket{VV}\bra{VV}.
\end{split}
\end{equation}
Experimentally, the expectation values of the correlation terms are obtained from coincidence counts $C$ measured in different projective bases, i.e.,
\begin{equation}\label{Eq:Expectation}
\begin{split}
\langle\sigma_x \otimes \sigma_x\rangle =& \frac{C_{++} - C_{+-} - C_{-+}+ C_{--}}{C_{++} + C_{+-} + C_{-+}+ C_{--}}, \\
\langle\sigma_y \otimes \sigma_y\rangle =& \frac{C_{LL} - C_{LR} - C_{RL}+ C_{RR}}{C_{LL} + C_{LR} + C_{RL}+ C_{RR}}, \\
\langle\sigma_z \otimes \sigma_z\rangle =& \frac{C_{HH} - C_{HV} - C_{VH}+ C_{VV}}{C_{HH} + C_{HV} + C_{VH}+ C_{VV}}.
\end{split}
\end{equation}

To implement the time-shift attack, we introduce a 5~ns timing offset on photon $v$ for selected measurement settings, while leaving photon $u$ unchanged. Since coincidence events are identified within a finite timing window, this offset causes a subset of genuine coincidence events to fall outside the window and be discarded. By applying the timing shift selectively, this effect becomes basis dependent. For a separable state $\sigma_{uv}$, we apply the time-shift attack to specific projection settings, i.e., $\ket{+-}\bra{+-}$, $\ket{-+}\bra{-+}$, $\ket{LL}\bra{LL}$, $\ket{RR}\bra{RR}$, $\ket{HV}\bra{HV}$ and $\ket{VH}\bra{VH}$, effectively suppressing the corresponding coincidence counts. According to Eq.~\ref{Eq:Expectation}, eliminating the contributions from $C_{+-}$, $C_{-+}$, $C_{LL}$, $C_{RR}$, $C_{HV}$ and $C_{VH}$ yields $\langle\sigma_x \sigma_x\rangle=-\langle\sigma_y \sigma_y\rangle=\langle\sigma_z \sigma_z\rangle=1$. Substituting these biased correlations into the witness expression (Eq.~\ref{Eq:EWsupp}) gives $\langle\mathcal W\rangle=-0.5$, causing a separable state $\sigma_{uv}$ to be falsely identified as entangled.

Conventional entanglement witness values for the separable state $\ket{e_ue_v}$ are measured for all six user pairs, both without and under a time-shift attack, as summarized in Table~\ref{tab:ew_attack_compare}.
\begin{table}[htbp]
\centering
\begin{tabular}{ccc}
\hline\hline
User pair & Without attack & Under attack \\
\hline\hline
$AB$ & $ +0.013 \pm 0.006$ & $-0.500 \pm 0.000$ \\
$AC$ & $ +0.013 \pm 0.005$ & $-0.500 \pm 0.000$ \\
$AD$ & $-0.012 \pm 0.007$ & $-0.500 \pm 0.000$ \\
$BC$ & $+0.019 \pm 0.007$ & $-0.499 \pm 0.001$ \\
$BD$ & $-0.007\pm 0.005$ & $-0.492 \pm 0.005$ \\
$CD$ & $+0.014 \pm 0.005 $ & $-0.495\pm 0.004$ \\
\hline\hline
\end{tabular}
\caption{Conventional entanglement witness values $\langle \mathcal{W} \rangle$ for the separable state $\ket{e_ue_v}$, measured for six user pairs without and under a time-shift attack.}
\label{tab:ew_attack_compare}
\end{table}

\section{Measurement-device-independent~(MDI) quantification of time-bin entangled photons}
\subsection{MDI entanglement witness}
In the MDI framework~\cite{branciard2013PRL}, the conventional entanglement witness $\mathcal W$ in Eq.~\ref{Eq:EWsupp} can be decomposed as a linear combination of trusted auxiliary input states
\begin{equation}
    \mathcal W=\sum_{s,t}\beta_{s,t}\,\tau_s^\top\otimes\omega_t^\top,
\end{equation}
where $\tau_s$ and $\omega_t$ denote the auxiliary states prepared by the two users, the superscript $\top$ indicates transposition in the computational basis, and $\beta_{s,t}$ are real coefficients.

For each choice of auxiliary inputs $\tau_s$ and $\omega_t$, the two distant users perform local joint measurements on the inputs together with their respective subsystems of the shared state $\rho_{uv}$. Writing $U_1$ and $V_1$ being their positive operator-valued measure~(POVM) elements corresponding to the outcomes 1, the probability that they both get this outcome is
\begin{equation}
    P(1,1|\tau_s,\omega_t)
    =
    \mathrm{Tr}\!\left[(U_1 \otimes V_1)(\tau_s \otimes \rho_{uv} \otimes \omega_t)\right].
\end{equation}
The associated MDI entanglement witness is then given by
\begin{equation}
    \mathcal I(\rho_{uv})=\sum_{s,t}\beta_{s,t}\,P(1,1|\tau_s,\omega_t).
\end{equation}

In the standard implementation, selecting $U_1=V_1=\ket{\Phi^+}\bra{\Phi^+}$, and the conditional probability simplifies to 
\begin{equation}
        P(1,1|\tau_s,\omega_t)
        = \mathrm{Tr}\!\left[(\ket{\Phi^+}\bra{\Phi^+}\otimes \ket{\Phi^+}\bra{\Phi^+})(\tau_s \otimes \rho_{uv} \otimes \omega_t)\right] = \frac{1}{4}\mathrm{Tr}\!\left[ (\tau_s^\top \otimes \omega_t^\top)\rho_{uv} \right],
\end{equation}
which leads to
\begin{equation}\label{Eq:EW-MDIEW}
        \mathcal I(\rho_{uv}) = \frac{1}{4}\sum_{s,t}\beta_{s,t}\,\mathrm{Tr}\!\left[ (\tau_s^\top \otimes \omega_t^\top)\rho_{uv} \right] = \frac{1}{4}\mathrm{Tr}\! \left (\mathcal W \rho_{uv} \right).
\end{equation}
This relation explicitly shows that $\mathcal I(\rho_{uv})$ reproduces the conventional witness value up to a constant factor, while depending only on observed conditional probabilities and trusted input states.

One convenient decomposition of the witness $\mathcal W=\sum_{s,t}\beta_{s,t}\,\tau_s^\top\otimes\omega_t^\top$ is obtained by choosing six pure auxiliary input states whose density matrices are
\begin{equation}
    \tau_s, \omega_t\in\{\ket{H}\bra{H}, \ket{V}\bra{V}, \ket{+}\bra{+}, \ket{-}\bra{-}, \ket{L}\bra{L}, \ket{R}\bra{R}\}.
\end{equation}
We use the same symbols $H,V,+,-,L,R$ to label the indices $s$ and $t$. 
With the rows and columns ordered as $(H,V,+,-,L,R)$, the coefficient matrix is
\begin{equation}
(\beta_{s,t}) =
\begin{pmatrix}
0 & \tfrac{1}{2} & 0 & 0 & 0 & 0\\
\tfrac{1}{2} & 0 & 0 & 0 & 0 & 0\\
0 & 0 & 0 & \tfrac{1}{2} & 0 & 0\\
0 & 0 & \tfrac{1}{2} & 0 & 0 & 0\\
0 & 0 & 0 & 0 & 0 & -\tfrac{1}{2}\\
0 & 0 & 0 & 0 & -\tfrac{1}{2} & 0
\end{pmatrix}.
\end{equation}
Equivalently, the nonzero coefficients are
\begin{equation}
    \beta_{H,V}=\beta_{V,H}
    =\beta_{+,-}=\beta_{-,+}
    =\frac{1}{2},
    \qquad
    \beta_{L,R}=\beta_{R,L}
    =-\frac{1}{2}.
\end{equation}
Since all other coefficients vanish, only these six pairs of auxiliary
input states contribute to the MDI witness. Thus, to reconstruct this
witness value experimentally, it is sufficient to measure the conditional probabilities associated with these six nonzero terms.
For the successful outcome pair $(1,1)$, the MDI witness value is
\begin{equation}
    \mathcal I(\rho_{uv})
    =
    \frac{1}{2}\big [P(1,1|\tau_H,\omega_V)
    +P(1,1|\tau_V,\omega_H)
    +P(1,1|\tau_+,\omega_-)
    +P(1,1|\tau_-,\omega_+)
    -P(1,1|\tau_L,\omega_R)
    -P(1,1|\tau_R,\omega_L) \big ].
\end{equation}    
For any separable state $\sigma_{uv}$, one has
$\mathcal I(\sigma_{uv})\ge 0,$
while a negative value $ \mathcal I(\rho_{uv})<0$ certifies entanglement in a MDI manner.

\subsection{MDI entanglement quantification}

In addition to entanglement detection, the MDI framework enables a quantitative characterization of entanglement directly from experimentally accessible data. In this section, we show how the measured MDI witness value can be used to derive a lower bound on a well-defined entanglement measure.

We consider the trace-distance measure of entanglement~\cite{vedral1997PRL}, defined as
\begin{equation}
    E_{\mathrm{Tr}}(\rho) = \min_{\sigma \in \Omega} D(\rho,\sigma),
\end{equation}
where $\Omega$ denotes the set of separable states and 
\begin{equation}
    D(\rho,\sigma) = \frac{1}{2}\mathrm{Tr}|\rho - \sigma|
\end{equation}
is the trace distance, with $\mathrm{Tr}|A| = \mathrm{Tr}\sqrt{AA^\dagger}$ the trace norm. Let $\sigma_{\mathrm{opt}} \in \Omega$ denote the optimal separable state achieving the minimum. The entanglement measure can then be written as
\begin{equation}
    E_{\mathrm{Tr}}(\rho) = \frac{1}{2}\mathrm{Tr}|\rho - \sigma_{\mathrm{opt}}|.
\end{equation}

In the MDI framework, the experimentally accessible quantities are the conditional probabilities
\begin{equation}
    P(1,1|\tau_s,\omega_t)
    = \mathrm{Tr}\!\left[(U_1 \otimes V_1)(\tau_s \otimes \rho \otimes \omega_t)\right].
\end{equation}
If the conventional entanglement witness admits the decomposition
\begin{equation}
    \mathcal W = \sum_{s,t} \beta_{s,t}\,\tau_s^\top \otimes \omega_t^\top,
\end{equation}
then the corresponding MDI witness value can be written as~\cite{sun2024PRL}
\begin{equation}
    \mathcal I(\rho) = \sum_{s,t}\beta_{s,t} P(1,1|\tau_s,\omega_t) = \mathrm{Tr}\!\left[(U_1 \otimes V_1)(\mathcal W^\top \otimes \rho)\right].
\end{equation}
This expression shows that the witness value can be directly evaluated from the experimentally measured probabilities.

Since $\sigma_{\mathrm{opt}}$ is separable, it satisfies the witness condition
\begin{equation}
    \mathcal I(\sigma_{\mathrm{opt}}) \ge 0.
\end{equation}
To relate the witness value to the entanglement measure, we use the identity
\begin{equation}
    \mathrm{Tr}|M \otimes N| = \mathrm{Tr}|M|\,\mathrm{Tr}|N|,
\end{equation}
valid for Hermitian operators $M$ and $N$. Setting $M = \mathcal W^\top$ and $N = \rho - \sigma_{\mathrm{opt}}$, we obtain
\begin{equation}
    2\,\mathrm{Tr}|\mathcal W^\top|\,E_{\mathrm{Tr}}(\rho)
    = \mathrm{Tr}|\mathcal W^\top \otimes (\rho - \sigma_{\mathrm{opt}})| 
    \ge \left| \mathrm{Tr}\!\left[(\mathcal W^\top \otimes (\rho - \sigma_{\mathrm{opt}}))(U_1 \otimes V_1)\right] \right|.
\end{equation}

Here we used the fact that for any operator $A$ and any operator $B$ satisfying $0 \le B \le I$, one has $\mathrm{Tr}|A| \ge |\mathrm{Tr}(AB)|$. This applies with $B = U_1 \otimes V_1$, since it is a POVM element. Evaluating the trace gives
\begin{equation}
    \mathrm{Tr}\!\left[(\mathcal W^\top \otimes (\rho - \sigma_{\mathrm{opt}}))(U_1 \otimes V_1)\right]
    = \mathcal I(\rho) - \mathcal I(\sigma_{\mathrm{opt}}),
\end{equation}
and therefore
\begin{equation}
    2\,\mathrm{Tr}|\mathcal W^\top|\,E_{\mathrm{Tr}}(\rho)
    \ge |\mathcal I(\rho) - \mathcal I(\sigma_{\mathrm{opt}})|.
\end{equation}
Using $\mathcal I(\sigma_{\mathrm{opt}}) \ge 0$, we obtain
\begin{equation}
    2\,\mathrm{Tr}|\mathcal W^\top|\,E_{\mathrm{Tr}}(\rho)
    \ge - \mathcal I(\rho),
\end{equation}
which leads to the final bound
\begin{equation}
    E_{\mathrm{Tr}}(\rho) \ge - \frac{\mathcal I(\rho)}{2\,\mathrm{Tr}|\mathcal W^\top|}.
\end{equation}
This result shows that a negative MDI witness value not only certifies entanglement, but also provides a quantitative lower bound on the trace-distance entanglement measure. Importantly, this bound is obtained directly from the experimentally measured conditional probabilities, without requiring additional measurements or any assumptions about the measurement devices. This establishes a direct operational link between MDI observations and quantitative entanglement characterization.

\subsection{Experimental setup for MDI measurement}

\begin{figure}[htb]
\centering
\includegraphics[width=0.5\textwidth]{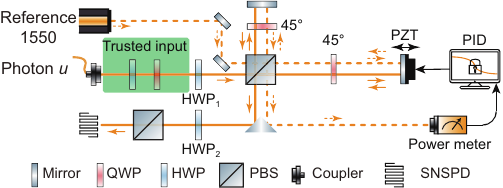}
\caption{The experimental setup to perform MDI measurements.}
\label{fig:MDI_setup}
\end{figure}

The experimental setup for implementing the MDI measurement closely resembles that of the projective measurement. In both cases, the measurement is based on the same UMZI architecture, with local phase stabilization and postselection of the middle time bin. The MDI measurement is realized with only minor modifications, as shown in Fig.~\ref{fig:MDI_setup}.
The MDI measurement implements a Bell-state measurement~(BSM) between the shared quantum state and the corresponding trusted input. In our experiment, this is realized as a hybrid BSM between the time-bin DoF~(carrying the shared state $\rho_{uv}$) and polarization DoF~(encoding the trusted input), within the Hilbert space spanned by $\{\ket{He},\ket{Ve},\ket{Hl},\ket{Vl}\}$. The four hybrid Bell states in this time-bin–polarization space are defined as
\begin{equation}
\begin{aligned}
\ket{\Phi^\pm} &= \frac{1}{\sqrt{2}}\left(\ket{He}\pm\ket{Vl}\right),\\
\ket{\Psi^\pm} &= \frac{1}{\sqrt{2}}\left(\ket{Hl}\pm\ket{Ve}\right).
\end{aligned}
\end{equation}

For an arbitrary state
\begin{equation}
\ket{\chi}=a\ket{He}+b\ket{Ve}+c\ket{Hl}+d\ket{Vl}, \quad |a|^2 + |b|^2 + |c|^2 + |d|^2 = 1,
\end{equation}
the corresponding Bell-state projection probabilities are
\begin{equation}
\begin{aligned}
P_{\Phi^+} &= |\braket{\Phi^+}{\chi}|^2 = \frac{|a+d|^2}{2},\\
P_{\Phi^-} &= |\braket{\Phi^-}{\chi}|^2 = \frac{|a-d|^2}{2},\\
P_{\Psi^+} &= |\braket{\Psi^+}{\chi}|^2 = \frac{|b+c|^2}{2},\\
P_{\Psi^-} &= |\braket{\Psi^-}{\chi}|^2 = \frac{|c-b|^2}{2}.
\end{aligned}
\end{equation}

Equivalently, the BSM can be summarized as
\begin{equation}
\ket{\chi}
\xrightarrow{\mathrm{BSM}}
\begin{cases}
\ket{\Phi^\pm} \Rightarrow P_{\Phi^\pm} = \dfrac{|a \pm d|^2}{2},\\[6pt]
\ket{\Psi^\pm} \Rightarrow P_{\Psi^\pm} = \dfrac{|b \pm c|^2}{2}.
\end{cases}
\end{equation}

To illustrate the operation of the BSM setup, we provide a step-by-step transformation in Eq.~\ref{eq:BSM}, using $\ket{\chi}$ as an example.
\begin{equation}
\begin{aligned}
\ket{\chi}
&\xrightarrow{\mathrm{HWP_1}}
\begin{cases}
\mathrm{HWP_1}@\:0^\circ\;\;\Rightarrow\;\; a\ket{He}-b\ket{Ve}+c\ket{Hl}-d\ket{Vl}\\[4pt]
\mathrm{HWP_1}@\:45^\circ \;\;\Rightarrow\;\; a\ket{Ve}+b\ket{He}+c\ket{Vl}+d\ket{Hl}
\end{cases}\\
&\xrightarrow{\mathrm{UMZI}}
\begin{cases}
\mathrm{HWP_1}@\:0^\circ\;\;\Rightarrow\;\; a e^{i\phi_u}\ket{Vm}-b\ket{He}+c e^{i\phi_u}\ket{Vl}-d\ket{Hm} \\[4pt]
\mathrm{HWP_1}@\:45^\circ \;\;\Rightarrow\;\; a\ket{He}+b e^{i\phi_u}\ket{Vm}+c\ket{Hm}+d e^{i\phi_u}\ket{Vl}
\end{cases}\\
&\xrightarrow[\phi_u=\pi]{\mathrm{HWP_2}}
\begin{cases}
\mathrm{HWP_1}@\:0^\circ,\;\mathrm{HWP_2}@\:22.5^\circ\;\;\Rightarrow\;\; \frac{d+a}{\sqrt{2}}\ket{Hm}+\frac{d-a}{\sqrt{2}}\ket{Vm}
+\frac{b}{\sqrt2}\big(\ket{He}+\ket{Ve}\big)
+\frac{c}{\sqrt2}\big(\ket{Hl}-\ket{Vl}\big) \\[4pt]
\mathrm{HWP_1}@\:45^\circ,\;\mathrm{HWP_2}@\:22.5^\circ \;\;\Rightarrow\;\; \frac{c-b}{\sqrt{2}}\ket{Hm}+\frac{b+c}{\sqrt{2}}\ket{Vm}
+\frac{a}{\sqrt2}\big(\ket{He}+\ket{Ve}\big)
-\frac{d}{\sqrt2}\big(\ket{Hl}-\ket{Vl}\big)\\[4pt]
\mathrm{HWP_1}@\:0^\circ,\;\mathrm{HWP_2}@\:67.5^\circ\;\;\Rightarrow\;\; \frac{a-d}{\sqrt{2}}\ket{Hm}+\frac{a+d}{\sqrt{2}}\ket{Vm}
-\frac{b}{\sqrt2}\big(\ket{He}-\ket{Ve}\big)
+\frac{c}{\sqrt2}\big(\ket{Hl}+\ket{Vl}\big) \\[4pt]
\mathrm{HWP_1}@\:45^\circ,\;\mathrm{HWP_2}@\:67.5^\circ \;\;\Rightarrow\;\; \frac{b+c}{\sqrt{2}}\ket{Hm}+\frac{b-c}{\sqrt{2}}\ket{Vm}
+\frac{a}{\sqrt2}\big(\ket{He}-\ket{Ve}\big)
+\frac{d}{\sqrt2}\big(\ket{Hl}+\ket{Vl}\big)
\end{cases}\\
&\xrightarrow[\mathrm{Postselection\,of\,}\ket{Hm}]{\mathrm{PBS}}
\begin{cases}
\ket{Hm}  \;\; \text{with probability of}\;\;\frac{|a+d|^2}{2}\;\;\Rightarrow\;\;\text{Projection\;on} \;\ket{\Phi^+}\\[4pt]
\ket{Hm} \;\; \text{with probability of}\;\;\frac{|c-b|^2}{2}\;\;\Rightarrow\;\;\text{Projection\;on} \;\ket{\Psi^-}\\[4pt]
\ket{Hm}  \;\; \text{with probability of}\;\;\frac{|a-d|^2}{2}\;\;\Rightarrow\;\;\text{Projection\;on} \;\ket{\Phi^-}\\[4pt]
\ket{Hm}  \;\; \text{with probability of}\;\;\frac{|b+c|^2}{2}\;\;\Rightarrow\;\;\text{Projection\;on} \;\ket{\Psi^+}\\[6pt]
\end{cases}
\end{aligned}
\label{eq:BSM}
\end{equation}
By postselecting the horizontally polarized middle time bin $\ket{Hm}$, the detection probabilities are given by $\frac{|a+d|^2}{2}, \frac{|c-b|^2}{2}, \frac{|a-d|^2}{2}$, and $\frac{|b+c|^2}{2}$, corresponding to projections onto the Bell states $\ket{\Phi^+}, \ket{\Psi^-}, \ket{\Phi^-}$ and $\ket{\Psi^+}$, respectively.

We perform MDI measurements on the six time-bin states $\ket{\Phi(\theta)}$ introduced in Sec.~\ref{fig:timebin_generate}. The MDI lower bounds and the corresponding trace-distance entanglement measures are summarized in Table~\ref{tab:benchmark_theta_scan}.

\begin{table}[htbp]
\centering
\begin{tabular}{cccc}
\hline\hline
Index & $\theta$ & MDI lower bound & $E_{\mathrm{Tr}}(\ket{\Phi(\theta)}\bra{\Phi(\theta)})$ \\
\hline\hline
1  & $0$          & $0.0017 \pm 0.0006$ & $0.010 \pm 0.012$\\
2  & $\pi/20$     & $0.0058 \pm 0.0008$ & $0.165 \pm 0.008$ \\
3  & $\pi/10$     & $0.0156 \pm 0.0009$ & $0.270 \pm 0.011$ \\
4  & $3\pi/20$    & $0.0212 \pm 0.0012$ & $0.328 \pm 0.011$\\
5  & $\pi/5$      & $0.0251 \pm 0.0010$ & $0.400 \pm 0.005$ \\
6  & $\pi/4$      & $0.0269 \pm 0.0012$ & $0.431 \pm 0.008$\\
\hline\hline
\end{tabular}
\caption{The MDI lower bound and the trace-distance measure of entanglement.}
\label{tab:benchmark_theta_scan}
\end{table}


\twocolumngrid
\bibliography{main}

@article{Wengerowsky2018Nature,
    author = {Wengerowsky, S{\"o}ren and Joshi, Siddarth Koduru and Steinlechner, Fabian and H{\"u}bel, Hannes and Ursin, Rupert},
    title = {An entanglement-based wavelength-multiplexed quantum communication network},
    journal = {Nature},
    year = {2018},
    volume = {564},
    number = {7735},
    pages = {225--228},
    month = dec,
    issn = {1476-4687},
    doi = {10.1038/s41586-018-0766-y},
    url = {https://doi.org/10.1038/s41586-018-0766-y}
}

@article{Huang2026NP,
    author = {Huang, Yiwen and Yang, Yilin and Li, Hao and Wang, Jiayu and Qiu, Jing and Qi, Zhantong and Zhang, Yuting and Li, Yuanhua and Zheng, Yuanlin and Chen, Xianfeng},
    title = {Quantum fusion of independent networks based on multi-user entanglement swapping},
    journal = {Nat. Photonics},
    year = {2026},
    volume = {20},
    number = {1},
    pages = {87--95},
    month = jan,
    issn = {1749-4893},
    doi = {10.1038/s41566-025-01792-0},
    url = {https://doi.org/10.1038/s41566-025-01792-0}
}

@article{Alshowkan2021PRXQ,
  title = {Reconfigurable Quantum Local Area Network Over Deployed Fiber},
  author = {Alshowkan, Muneer and Williams, Brian P. and Evans, Philip G. and Rao, Nageswara S.V. and Simmerman, Emma M. and Lu, Hsuan-Hao and Lingaraju, Navin B. and Weiner, Andrew M. and Marvinney, Claire E. and Pai, Yun-Yi and Lawrie, Benjamin J. and Peters, Nicholas A. and Lukens, Joseph M.},
  journal = {PRX Quantum},
  volume = {2},
  issue = {4},
  pages = {040304},
  numpages = {13},
  year = {2021},
  month = {Oct},
  publisher = {American Physical Society},
  doi = {10.1103/PRXQuantum.2.040304},
  url = {https://link.aps.org/doi/10.1103/PRXQuantum.2.040304}
}

@article{Joshi2020Sci.Adv.,
author = {Siddarth Koduru Joshi  and Djeylan Aktas  and Sören Wengerowsky  and Martin Lončarić  and Sebastian Philipp Neumann  and Bo Liu  and Thomas Scheidl  and Guillermo Currás Lorenzo  and Željko Samec  and Laurent Kling  and Alex Qiu  and Mohsen Razavi  and Mario Stipčević  and John G. Rarity  and Rupert Ursin },
title = {A trusted node–free eight-user metropolitan quantum communication network},
journal = {Sci. Adv.},
volume = {6},
number = {36},
pages = {eaba0959},
year = {2020},
doi = {10.1126/sciadv.aba0959},
URL = {https://www.science.org/doi/abs/10.1126/sciadv.aba0959}}

@article{Shi2025arXiv,
      title={Integrated polarization-entangled photon source for wavelength-multiplexed quantum networks}, 
      author={Xiaodong Shi and Yue Li and Jinyi Du and Lin Zhou and Ran Yang and En Teng Lim and Sakthi Sanjeev Mohanraj and Mengyao Zhao and Xu Chen and Xiaojie Wang and Guangxing Wu and Hao Hao and Veerendra Dhyani and Sihao Wang and Alexander Ling and Di Zhu},
      year={2025},
      journal={arXiv:2511.22680},
      archivePrefix={arXiv},
      primaryClass={physics.optics},
      url={https://arxiv.org/abs/2511.22680}, 
}

@article{Huang2025LPR,
    author = {Huang, Yiwen and Qi, Zhantong and Yang, Yilin and Zhang, Yuting and Li, Yuanhua and Zheng, Yuanlin and Chen, Xianfeng},
    title = {A Sixteen-user Time-bin Entangled Quantum Communication Network With Fully Connected Topology},
    journal = {Laser Photonics Rev.},
    year = {2025},
    volume = {19},
    number = {1},
    pages = {2301026},
    issn = {1863-8880},
    doi = {10.1002/lpor.202301026},
    url = {https://doi.org/10.1002/lpor.202301026}
}

@article{Kim2022APL.Photon.,
    author = {Kim, Jin-Hun and Chae, Jin-Woo and Jeong, Youn-Chang and Kim, Yoon-Ho},
    title = {Quantum communication with time-bin entanglement over a wavelength-multiplexed fiber network},
    journal = {APL Photonics},
    volume = {7},
    number = {1},
    pages = {016106},
    year = {2022},
    month = {01},
    issn = {2378-0967},
    doi = {10.1063/5.0073040},
    url = {https://doi.org/10.1063/5.0073040}
}

@article{Fitzke2022PRXQ,
  title = {Scalable Network for Simultaneous Pairwise Quantum Key Distribution via Entanglement-Based Time-Bin Coding},
  author = {Fitzke, Erik and Bialowons, Lucas and Dolejsky, Till and Tippmann, Maximilian and Nikiforov, Oleg and Walther, Thomas and Wissel, Felix and Gunkel, Matthias},
  journal = {PRX Quantum},
  volume = {3},
  issue = {2},
  pages = {020341},
  numpages = {14},
  year = {2022},
  month = {May},
  publisher = {American Physical Society},
  doi = {10.1103/PRXQuantum.3.020341},
  url = {https://link.aps.org/doi/10.1103/PRXQuantum.3.020341}
}

@article{Liu2022PhotoniX,
    author = {Liu, Xu and Liu, Jingyuan and Xue, Rong and Wang, Heqing and Li, Hao and Feng, Xue and Liu, Fang and Cui, Kaiyu and Wang, Zhen and You, Lixing and Huang, Yidong and Zhang, Wei},
    title = {40-user fully connected entanglement-based quantum key distribution network without trusted node},
    journal = {PhotoniX},
    year = {2022},
    volume = {3},
    number = {1},
    pages = {2},
    month = jan,
    issn = {2662-1991},
    doi = {10.1186/s43074-022-00048-2},
    url = {https://doi.org/10.1186/s43074-022-00048-2}
}

@article{Li2019PRL,
  title = {Multiuser Time-Energy Entanglement Swapping Based on Dense Wavelength Division Multiplexed and Sum-Frequency Generation},
  author = {Li, Yuanhua and Huang, Yiwen and Xiang, Tong and Nie, Yiyou and Sang, Minghuang and Yuan, Luqi and Chen, Xianfeng},
  journal = {Phys. Rev. Lett.},
  volume = {123},
  issue = {25},
  pages = {250505},
  numpages = {5},
  year = {2019},
  month = {Dec},
  publisher = {American Physical Society},
  doi = {10.1103/PhysRevLett.123.250505},
  url = {https://link.aps.org/doi/10.1103/PhysRevLett.123.250505}
}

@article{Xavier2025npjQI,
    author = {Xavier, Guilherme B. and Larsson, Jan-{\AA}ke and Villoresi, Paolo and Vallone, Giuseppe and Cabello, Ad{\'a}n},
    title = {Energy-time and time-bin entanglement: past, present and future},
    journal = { npj Quantum Inf.},
    year = {2025},
    volume = {11},
    number = {1},
    pages = {129},
    month = jul,
    issn = {2056-6387},
    doi = {10.1038/s41534-025-01072-3},
    url = {https://doi.org/10.1038/s41534-025-01072-3}
}

@article{james2001PRA,
  title = {Measurement of Qubits},
  author = {James, Daniel F. V. and Kwiat, Paul G. and Munro, William J. and White, Andrew G.},
  year = 2001,
  month = oct,
  journal = {Phys. Rev. A},
  volume = {64},
  number = {5},
  pages = {052312},
  publisher = {American Physical Society},
  doi = {10.1103/PhysRevA.64.052312}
}

@article{thew2002PRA,
  title = {Qudit Quantum-State Tomography},
  author = {Thew, R. T. and Nemoto, K. and White, A. G. and Munro, W. J.},
  year = 2002,
  month = jul,
  journal = {Phys. Rev. A},
  volume = {66},
  number = {1},
  pages = {012303},
  publisher = {American Physical Society},
  doi = {10.1103/PhysRevA.66.012303}
}

@article{horodecki1996PLA,
  title = {Separability of Mixed States: Necessary and Sufficient Conditions},
  shorttitle = {Separability of Mixed States},
  author = {Horodecki, Micha{\l} and Horodecki, Pawe{\l} and Horodecki, Ryszard},
  year = 1996,
  month = nov,
  journal = {Phys. Lett. A},
  volume = {223},
  number = {1-2},
  pages = {1--8},
  issn = {03759601},
  doi = {10.1016/S0375-9601(96)00706-2}
}

@article{terhal2000PLA,
  title = {Bell Inequalities and the Separability Criterion},
  author = {Terhal, Barbara M.},
  year = 2000,
  month = jul,
  journal = {Phys. Lett. A},
  volume = {271},
  number = {5-6},
  pages = {319--326},
  issn = {03759601},
  doi = {10.1016/S0375-9601(00)00401-1}
}

@article{guhne2009PR,
  title = {Entanglement Detection},
  author = {G{\"u}hne, Otfried and T{\'o}th, G{\'e}za},
  year = 2009,
  month = apr,
  journal = {Phys. Rep.},
  volume = {474},
  number = {1-6},
  pages = {1--75},
  issn = {03701573},
  doi = {10.1016/j.physrep.2009.02.004}
}

@article{rosset2012PRA,
  title = {Imperfect Measurement Settings: {{Implications}} for Quantum State Tomography and Entanglement Witnesses},
  shorttitle = {Imperfect Measurement Settings},
  author = {Rosset, Denis and {Ferretti-Sch{\"o}bitz}, Raphael and Bancal, Jean-Daniel and Gisin, Nicolas and Liang, Yeong-Cherng},
  year = 2012,
  month = dec,
  journal = {Phys. Rev. A},
  volume = {86},
  number = {6},
  pages = {062325},
  issn = {1050-2947, 1094-1622},
  doi = {10.1103/PhysRevA.86.062325}
}

@article{buscemi2012PRL,
  title = {All Entangled Quantum States Are Nonlocal},
  author = {Buscemi, Francesco},
  year = 2012,
  month = may,
  journal = {Phys. Rev. Lett.},
  volume = {108},
  number = {20},
  pages = {200401},
  issn = {0031-9007, 1079-7114},
  doi = {10.1103/PhysRevLett.108.200401}
}

@article{branciard2013PRL,
  title = {Measurement-Device-Independent Entanglement Witnesses for All Entangled Quantum States},
  author = {Branciard, Cyril and Rosset, Denis and Liang, Yeong-Cherng and Gisin, Nicolas},
  year = 2013,
  month = feb,
  journal = {Phys. Rev. Lett.},
  volume = {110},
  number = {6},
  pages = {060405},
  issn = {0031-9007, 1079-7114},
  doi = {10.1103/PhysRevLett.110.060405}
}

@article{xu2014PRL,
  title = {Implementation of a Measurement-Device-Independent Entanglement Witness},
  author = {Xu, Ping and Yuan, Xiao and Chen, Luo-Kan and Lu, He and Yao, Xing-Can and Ma, Xiongfeng and Chen, Yu-Ao and Pan, Jian-Wei},
  year = 2014,
  month = apr,
  journal = {Phys. Rev. Lett.},
  volume = {112},
  number = {14},
  pages = {140506},
  issn = {0031-9007, 1079-7114},
  doi = {10.1103/PhysRevLett.112.140506}
}

@article{verbanis2016PRL,
  title = {Resource-Efficient Measurement-Device-Independent Entanglement Witness},
  author = {Verbanis, E. and Martin, A. and Rosset, D. and Lim, C. C. W. and Thew, R. T. and Zbinden, H.},
  year = 2016,
  month = may,
  journal = {Phys. Rev. Lett.},
  volume = {116},
  number = {19},
  pages = {190501},
  issn = {0031-9007, 1079-7114},
  doi = {10.1103/PhysRevLett.116.190501}
}

@article{li2020PRL,
  title = {Measurement-Device-Independent Entanglement Witness of Tripartite Entangled States and Its Applications},
  author = {Li, Zheng-Da and Zhao, Qi and Zhang, Rui and Liu, Li-Zheng and Yin, Xu-Fei and Zhang, Xingjian and Fei, Yue-Yang and Chen, Kai and Liu, Nai-Le and Xu, Feihu and Chen, Yu-Ao and Li, Li and Pan, Jian-Wei},
  year = 2020,
  month = apr,
  journal = {Phys. Rev. Lett.},
  volume = {124},
  number = {16},
  pages = {160503},
  issn = {0031-9007, 1079-7114},
  doi = {10.1103/PhysRevLett.124.160503}
}

@article{abiuso2021PRL,
  title = {Measurement-Device-Independent Entanglement Detection for Continuous-Variable Systems},
  author = {Abiuso, Paolo and B{\"a}uml, Stefan and Cavalcanti, Daniel and Ac{\'i}n, Antonio},
  year = 2021,
  month = may,
  journal = {Phys. Rev. Lett.},
  volume = {126},
  number = {19},
  pages = {190502},
  issn = {0031-9007, 1079-7114},
  doi = {10.1103/PhysRevLett.126.190502}
}

@article{wang2025PRL,
  title = {Experimental Measurement-Device-Independent Verification of Continuous-Variable Entanglement},
  author = {Wang, Xutong and Fu, Jing and Jing, Jietai},
  year = 2025,
  month = oct,
  journal = {Phys. Rev. Lett.},
  volume = {135},
  number = {15},
  pages = {150201},
  issn = {0031-9007, 1079-7114},
  doi = {10.1103/rk3t-69tb}
}

@article{fu2025LSA,
  title = {Measurement-Device-Independent Continuous Variable Entanglement Witness in a Quantum Network},
  author = {Fu, Jing and Wang, Xutong and Liu, Shengshuai and Jing, Jietai},
  year = 2025,
  month = nov,
  journal = {Light Sci. Appl.},
  volume = {14},
  number = {1},
  pages = {376},
  issn = {2047-7538},
  doi = {10.1038/s41377-025-02039-x}
}

@article{vedral1997PRL,
  title = {Quantifying Entanglement},
  author = {Vedral, V. and Plenio, M. B. and Rippin, M. A. and Knight, P. L.},
  year = 1997,
  month = mar,
  journal = {Phys. Rev. Lett.},
  volume = {78},
  number = {12},
  pages = {2275--2279},
  issn = {0031-9007, 1079-7114},
  doi = {10.1103/PhysRevLett.78.2275}
}

@article{shahandeh2017PRL,
  title = {Measurement-Device-Independent Approach to Entanglement Measures},
  author = {Shahandeh, Farid and Hall, Michael J. W. and Ralph, Timothy C.},
  year = 2017,
  month = apr,
  journal = {Phys. Rev. Lett.},
  volume = {118},
  number = {15},
  pages = {150505},
  publisher = {American Physical Society},
  doi = {10.1103/PhysRevLett.118.150505}
}

@article{sun2024PRL,
  title = {Bounding the Amount of Entanglement from Witness Operators},
  author = {Sun, Liang-Liang and Zhou, Xiang and Tavakoli, Armin and Xu, Zhen-Peng and Yu, Sixia},
  year = 2024,
  month = mar,
  journal = {Phys. Rev. Lett.},
  volume = {132},
  number = {11},
  pages = {110204},
  issn = {0031-9007, 1079-7114},
  doi = {10.1103/PhysRevLett.132.110204},
}

@article{sarkar2026NP,
  title = {A Universal Scheme to Self-Test Any Quantum State or Measurement},
  author = {Sarkar, Shubhayan and Orthey Jr, Alexandre C. and Augusiak, Remigiusz},
  year = 2026,
  month = mar,
  journal = {Nat. Phys.},
  volume = {22},
  number = {3},
  pages = {446--451},
  issn = {1745-2481},
  doi = {10.1038/s41567-026-03181-y}
}

@article{morelli2022PRL,
  title = {Entanglement Detection with Imprecise Measurements},
  author = {Morelli, Simon and Yamasaki, Hayata and Huber, Marcus and Tavakoli, Armin},
  year = 2022,
  month = jun,
  journal = {Phys. Rev. Lett.},
  volume = {128},
  number = {25},
  pages = {250501},
  publisher = {American Physical Society},
  doi = {10.1103/PhysRevLett.128.250501}
}

@article{cao2024PRL,
  title = {Genuine Multipartite Entanglement Detection with Imperfect Measurements: {{Concept}} and Experiment},
  author = {Cao, Huan and Morelli, Simon and Rozema, Lee A. and Zhang, Chao and Tavakoli, Armin and Walther, Philip},
  year = 2024,
  month = oct,
  journal = {Phys. Rev. Lett.},
  volume = {133},
  number = {15},
  pages = {150201},
  publisher = {American Physical Society},
  doi = {10.1103/PhysRevLett.133.150201}
}

@article{Lo2023QST,
doi = {10.1088/2058-9565/acc7c2},
url = {https://doi.org/10.1088/2058-9565/acc7c2},
year = {2023},
month = {apr},
publisher = {IOP Publishing},
volume = {8},
number = {3},
pages = {035003},
author = {Lo, Hsin-Pin and Ikuta, Takuya and Azuma, Koji and Honjo, Toshimori and Munro, William J and Takesue, Hiroki},
title = {Generation of a time–bin Greenberger–Horne–Zeilinger state with an optical switch},
journal = {Quantum Sci. Technol.}
}

@article{Reimer2016Science,
author = {Christian Reimer  and Michael Kues  and Piotr Roztocki  and Benjamin Wetzel  and Fabio Grazioso  and Brent E. Little  and Sai T. Chu  and Tudor Johnston  and Yaron Bromberg  and Lucia Caspani  and David J. Moss  and Roberto Morandotti },
title = {Generation of multiphoton entangled quantum states by means of integrated frequency combs},
journal = {Science},
volume = {351},
number = {6278},
pages = {1176-1180},
year = {2016},
doi = {10.1126/science.aad8532},
URL = {https://www.science.org/doi/abs/10.1126/science.aad8532}
}

@article{fang2026arXiv,
      title={Chip-Integrated Broadband Multi-Photon Source for Wavelength-Multiplexed Quantum Networks}, 
      author={Xiao-Xu Fang and Ling-Xuan Kong and He Lu},
      year={2026},
      journal = {arXiv:2603.09397}, 
      url={https://arxiv.org/abs/2603.09397}, 
}

@article{Horodecki1999PRA,
  title = {Reduction criterion of separability and limits for a class of distillation protocols},
  author = {Horodecki, Michal and Horodecki, Pawel},
  journal = {Phys. Rev. A},
  volume = {59},
  pages = {4206--4216},
  year = {1999},
  doi = {10.1103/PhysRevA.59.4206}
}

@article{zhao2016PRA,
  title = {Efficient measurement-device-independent detection of multipartite entanglement structure},
  author = {Zhao, Qi and Yuan, Xiao and Ma, Xiongfeng},
  journal = {Phys. Rev. A},
  volume = {94},
  issue = {1},
  pages = {012343},
  numpages = {8},
  year = {2016},
  month = {Jul},
  publisher = {American Physical Society},
  doi = {10.1103/PhysRevA.94.012343},
  url = {https://link.aps.org/doi/10.1103/PhysRevA.94.012343}
}

@article{Terhal2000PRA,
  title = {Schmidt number for density matrices},
  author = {Terhal, Barbara M. and Horodecki, Pawe\l{}},
  journal = {Phys. Rev. A},
  volume = {61},
  issue = {4},
  pages = {040301},
  numpages = {4},
  year = {2000},
  month = {Mar},
  publisher = {American Physical Society},
  doi = {10.1103/PhysRevA.61.040301},
  url = {https://link.aps.org/doi/10.1103/PhysRevA.61.040301}
}

@article{javid2021PRL,
  ids = {javid2021_Phys.Rev.Lett._Ultrabroadbanda},
  title = {Ultrabroadband {{Entangled Photons}} on a {{Nanophotonic Chip}}},
  author = {Javid, Usman A. and Ling, Jingwei and Staffa, Jeremy and Li, Mingxiao and He, Yang and Lin, Qiang},
  year = 2021,
  month = oct,
  journal = {Phys. Rev. Lett.},
  volume = {127},
  number = {18},
  pages = {183601},
  issn = {0031-9007, 1079-7114},
  doi = {10.1103/PhysRevLett.127.183601}
}

@article{fang2026OEO,
  title = {Broadband Quantum Photon Source in a Step-Chirped Periodically Poled Lithium Niobate Waveguide},
  author = {Fang, Xiao-Xu and Shentu, Guoliang and Lu, He},
  date = {2026-02-09},
  year = {2026}, 
  journal = {Opt. Express},
  volume = {34},
  number = {3},
  pages = {3759--3767},
  publisher = {Optica Publishing Group},
  issn = {1094-4087},
  doi = {10.1364/OE.582490},
  url = {https://opg.optica.org/oe/abstract.cfm?uri=oe-34-3-3759}
}

\end{document}